\input harvmac
\input psfig
\input amssym.def
\input amssym.tex

\font\ensX=msbm10
\font\ensVII=msbm7
\font\ensV=msbm5
\newfam\math
\textfont\math=\ensX \scriptfont\math=\ensVII \scriptscriptfont\math=\ensV
\def\ensemble{\fam\math\ensX}
\def\half{{1\over2}}
\def\IR{{\ensemble R}}
\def\wed{\mathbin{\scriptstyle \mathchar"225E}}
\def\Id{{\ensemble I}}

\def\Half(#1){\textstyle{#1\over2}}

\def\Box#1{\mathop{\mkern0.5\thinmuskip
           \vbox{\hrule\hbox{\vrule\hskip#1\vrule height#1 width 0pt\vrule}
           \hrule}\mkern0.5\thinmuskip}}

\noblackbox

\lref\VerlindeTX{E.~Verlinde and H.~Verlinde, {\sl Lectures On String
Perturbation Theory,} IASSNS-HEP-88-52
{\it Presented at Spring School and Workshop on Superstrings, Trieste, Italy, April 11-22, 1988}
}
\lref\deWitNM{B.~de Wit and A.~Van Proeyen,
{\sl Special geometry, cubic polynomials and homogeneous quaternionic spaces,}
Commun.\ Math.\ Phys.\  {\bf 149}, 307 (1992)
[arXiv:hep-th/9112027].
}

\lref\CecottiAD{S.~Cecotti,
{\sl Homogeneous Kahler Manifolds And T Algebras In N=2 Supergravity And
Superstrings,} Commun.\ Math.\ Phys.\  {\bf 124}, 23 (1989).
}

\lref\WessCP{J.~Wess and J.~Bagger, {\sl Supersymmetry And Supergravity}, Princeton, USA: Univ. Pr. (1992)
}

\lref\VafaWitten{C.~Vafa and E.~Witten, {\sl A One loop test of string
duality}, Nucl.\ Phys.\ B {\bf 447}, 261 (1995), [arXiv:hep-th/9505053].
}

\lref\JDuffTLiuMinasian{M.~J.~Duff,
J.~T.~Liu and R.~Minasian, {\sl Eleven-dimensional origin of string /
string duality: A one-loop test}, Nucl.\ Phys.\ B {\bf 452}, 261 (1995),
[arXiv:hep-th/9506126].
}

\lref\Minahan{J.~A.~Minahan,
{\sl One Loop Amplitudes On Orbifolds And The Renormalization Of Coupling
Constants,}  Nucl.\ Phys.\ B {\bf 298}, 36 (1988).
}

\lref\ForgerVJ{K.~Forger, B.~A.~Ovrut, S.~J.~Theisen and D.~Waldram,
{\sl Higher-Derivative Gravity in String Theory,} Phys.\ Lett.\ B {\bf 388},
512 (1996) [arXiv:hep-th/9605145].
}

\lref\DeJaegherKA{
J.~De Jaegher, B.~de Wit, B.~Kleijn and S.~Vandoren, {\sl Special geometry in hypermultiplets,} 
Nucl.\ Phys.\ B {\bf 514}, 553 (1998)
[arXiv:hep-th/9707262].
}

\lref\PiolineMN{B.~Pioline, {\sl A note on non-perturbative $R^4$ couplings,} 
Phys.\ Lett.\ B {\bf 431}, 73 (1998) [arXiv:hep-th/9804023].
}

\lref\GreenBY{M.~B.~Green and S.~Sethi, {\sl Supersymmetry constraints on 
type IIB supergravity,}
Phys.\ Rev.\ D {\bf 59}, 046006 (1999) [arXiv:hep-th/9808061].
}

\lref\KehagiasCQ{
A.~Kehagias and H.~Partouche,
{\sl The exact quartic effective action for the type IIB superstring,}
Phys.\ Lett.\ B {\bf 422}, 109 (1998)
[arXiv:hep-th/9710023].
}

\lref\KetovVR{
S.~V.~Ketov, {\sl Summing up D-instantons in N = 2 supergravity,}
Nucl.\ Phys.\ B {\bf 649}, 365 (2003)
[arXiv:hep-th/0209003].
}

\lref\FreemanZH{
M.~D.~Freeman, C.~N.~Pope, M.~F.~Sohnius and K.~S.~Stelle,
{\sl Higher Order Sigma Model Counterterms And The Effective Action
For Superstrings,} Phys.\ Lett.\ B {\bf 178}, 199 (1986).
}

\lref\KasteMV{P. Kaste, R. Minasian and P. Vanhove, Work in Progress.}

\lref\CadavidBK{A.~C.~Cadavid, A.~Ceresole, R.~D'Auria and S.~Ferrara, {\sl
Eleven-dimensional supergravity compactified on Calabi-Yau threefolds,}
Phys.\ Lett.\ B {\bf 357}, 76 (1995)
[arXiv:hep-th/9506144].
}

\lref\GreenTV{M.~B.~Green and M.~Gutperle, {\sl Effects of D-instantons,}
Nucl.\ Phys.\ B {\bf 498}, 195 (1997) [arXiv:hep-th/9701093].
}
\lref\BeckerNN{K.~Becker, M.~Becker, M.~Haack and J.~Louis,
{\sl Supersymmetry breaking and $\alpha'$ corrections to flux induced
potentials,} JHEP {\bf 0206}, 060 (2002) [arXiv:hep-th/0204254].
}

\lref\GreenAS{M.~B  Green, M.~Gutperle and P.~Vanhove, {\sl One loop in eleven
dimensions,} Phys.\ Lett.\ B {\bf 409}, 177 (1997) [arXiv:hep-th/9706175].
}

\lref\PeetersVW{
K.~Peeters, P.~Vanhove and A.~Westerberg,
{\sl Supersymmetric higher-derivative actions in ten and eleven dimensions,  the associated superalgebras 
and their formulation in superspace,}
Class.\ Quant.\ Grav.\  {\bf 18}, 843 (2001)
[arXiv:hep-th/0010167].
}

\lref\BerkovitsCB{N.~Berkovits and W.~Siegel, {\sl Superspace Effective Actions for 4D Compactifications of 
Heterotic and Type II Superstrings,}
Nucl.\ Phys.\ B {\bf 462}, 213 (1996) [arXiv:hep-th/9510106].
}

\lref\WittenEX{E.~Witten, {\sl String theory dynamics in various dimensions,}
Nucl.\ Phys.\ B {\bf 443}, 85 (1995) [arXiv:hep-th/9503124].
}

\lref\GuntherSC{H.~Gunther, C.~Herrmann and J.~Louis, 
{\sl Quantum corrections in the hypermultiplet moduli space}, Fortsch.\ Phys.\  {\bf 48}, 119 (2000) 
[arXiv:hep-th/9901137]. 
}

\lref\StromingerEB{A.~Strominger, 
{\sl Loop corrections to the universal hypermultiplet}, Phys.\ Lett.\ B {\bf 421}, 139 (1998) 
[arXiv:hep-th/9706195]. 
}

\lref\BaggerJF{J.~Bagger, {\sl Hypermultiplet Couplings In N=2
Supergravity,} SLAC-PUB-3405 {\it Invited talk at the Int. Conf. on High Energy Physics, Leipzig, 
East Germany, Jul 19-25, 1984}
}

\lref\rfCP{D.M.J. Calderbank and H. Pedersen, {\sl Selfdual Einstein metrics
with torus symmetry}, J. Diff. Geom. {\bf 60} (2002) 485-521, [arXiv:math.DG/0105263];\hfil\break
David M. J. Calderbank, Michael A. Singer, {\sl Einstein metrics and complex singularities},   
[arXiv:math.DG/0206229].}

\lref\AntoniadisEG{I.~Antoniadis, S.~Ferrara, R.~Minasian and K.~S.~Narain, 
{\sl $R^4$ couplings in M- and type II theories on Calabi-Yau spaces}, 
Nucl.\ Phys.\ B {\bf 507}, 571 (1997) [arXiv:hep-th/9707013]. 
}
\lref\AntoniadisTR{I.~Antoniadis, R.~Minasian and P.~Vanhove, 
{\sl Non-compact Calabi-Yau manifolds and localized gravity}, Nucl.\ Phys.\ B {\bf 648}, 69 (2003) 
[arXiv:hep-th/0209030]. 
}

\lref\PeetersUB{K.~Peeters, P.~Vanhove and A.~Westerberg, 
{\sl Chiral splitting and world-sheet gravitinos in higher-derivative string
amplitudes}, Class.\ Quant.\ Grav.\  {\bf 19}, 2699 (2002)
[arXiv:hep-th/0112157].  
}

\lref\BaggerTT{J.~Bagger and E.~Witten, {\sl Matter Couplings In N=2
Supergravity,} Nucl.\ Phys.\ B {\bf 222}, 1 (1983).
}

\lref\ZachosIW{C.~K.~Zachos, {\sl N=2 Supergravity Theory With A Gauged
Central Charge,} Phys.\ Lett.\ B {\bf 76}, 329 (1978).
}

\lref\FerraraIK{S.~Ferrara and S.~Sabharwal,
{\sl Quaternionic Manifolds For Type II Superstring Vacua Of
Calabi-Yau Spaces,} Nucl.\ Phys.\ B {\bf 332}, 317 (1990). 
}  

\lref\FerraraFF{S.~Ferrara and S.~Sabharwal, {\sl Dimensional Reduction Of
Type II Superstrings,} Class.\ Quant.\ Grav.\  {\bf 6}, L77 (1989).
}

\lref\BodnerCG{M.~Bodner and A.~C.~Cadavid, {\sl Dimensional Reduction Of Type
IIb Supergravity And Exceptional Quaternionic Manifolds,} Class.\ Quant.\
Grav.\  {\bf 7}, 829 (1990). 
}

\lref\AtickRS{J.~J.~Atick and A.~Sen, {\sl Covariant One Loop Fermion Emission
Amplitudes In Closed String Theories,} Nucl.\ Phys.\ B {\bf 293}, 317 (1987).
}

\lref\PasquinucciIK{A.~Pasquinucci and K.~Roland,
{\sl On the computation of one loop amplitudes with external fermions in 4-D
heterotic superstrings,} Nucl.\ Phys.\ B {\bf 440}, 441 (1995)
[arXiv:hep-th/9411015].
}

\lref\CecottiQN{S.~Cecotti, S.~Ferrara and L.~Girardello,
{\sl Geometry Of Type II Superstrings And The Moduli Of Superconformal Field
 Theories,} Int.\ J.\ Mod.\ Phys.\ A {\bf 4}, 2475 (1989).
}

\lref\GrossMW{D.~J.~Gross and J.~H.~Sloan, {\sl The Quartic Effective Action
For The Heterotic String,} Nucl.\ Phys.\ B {\bf 291}, 41 (1987).
}

\lref\BernPK{Z.~Bern and D.~A.~Kosower, {\sl Absence Of Wave Function Renormalization In Polyakov Amplitudes,}
Nucl.\ Phys.\ B {\bf 321}, 605 (1989).
}
\noblackbox
\baselineskip 14pt plus 2pt minus 2pt
\Title{\vbox{\baselineskip12pt
\hbox{hep-th/0307268}
\hbox{AEI-2003-052}
\hbox{CPHT-RR 030.0603}
\hbox{CERN-TH/2003-139}
\hbox{SPHT-T03/088}
}}
{\vbox{
\centerline{String loop corrections to the universal hypermultiplet}
}}
\centerline{Ignatios Antoniadis$^{1,a,}$\footnote{${}^\dagger$}{\sevenrm On leave of absence 
from CPHT {\'E}cole Polytechnique$^{2}$.}, Ruben Minasian$^{2,b}$, Stefan Theisen $^{3,c}$ and Pierre Vanhove$^{1,4,d}$}
\bigskip
\centerline{${}^1$ \sl CERN Theory Division CH-1211 Geneva 23, Switzerland}
\centerline{${}^2$ \sl CPHT {\'E}cole Polytechnique (UMR du CNRS 7644)
F-91128, Palaiseau}
\centerline{${}^3$ \sl
Max-Planck-Institut f{\"u}r  Gravitationsphysik,
Albert-Einstein-Institut, D-14476 Golm}
\centerline{${}^4$ \sl CEA/DSM/SPhT, URA 
au CNRS, CEA/Saclay, F-91191 Gif-sur-Yvette}
\bigskip
\centerline{$^a${\tt ignatios.antoniadis@cern.ch},
$^b${\tt ruben@cpht.polytechnique.fr}}
\centerline{
$^c${\tt theisen@aei.mpg.de},
$^d${\tt vanhove@spht.saclay.cea.fr}}
\bigskip

\bigskip
\centerline{{\bf Abstract}}
\medskip
We study loop corrections to the universal dilaton supermultiplet for type IIA
strings compactified on Calabi-Yau threefolds. We show that the
corresponding quaternionic kinetic terms receive non-trivial one-loop
contributions proportional to the Euler number of the Calabi-Yau
manifold, while the higher-loop corrections can be absorbed by field
redefinitions. The corrected metric is no longer K{\"a}hler. Our analysis
implies in particular that the Calabi-Yau volume is renormalized by loop
effects which are present even in higher orders, while there are also
one-loop corrections to the Bianchi identities for the NS and RR field
strengths.

\noblackbox
\baselineskip 14pt plus 2pt minus 2pt
\Date{}

\newsec{Introduction and discussion}
Type II string compactifications on Calabi-Yau threefolds (CY$_3$)
provide a theoretical framework for addressing several physically
interesting problems. Away from possible brane insertions, the
four-dimensional (4d) low energy massless spectrum is ${\cal N}=2$
supersymmetric and has two separate and decoupled matter sectors
involving, respectively, the vectors and hypermultiplets. At a generic
point of the moduli space, the vectors are abelian and the hypermultiplets are
neutral, while their corresponding multiplicities are given by the Betti
numbers of the $(1,1)$ and $(1,2)$ forms of the CY$_3$: $h_{(1,1)}$
($h_{(1,2)}$) and $h_{(1,2)}+1$ ($h_{(1,1)}+1$) in type IIA (IIB) theory.
The $+1$ stands for the so-called universal hypermultiplet,
formed by the 4d dilaton, the axion dual to the NS-NS (Neveu-Schwarz--Neveu-Schwarz)
2-form and a complex RR (Ramond-Ramond) scalar $C$, obtained for instance
in type IIA, by the 3-form gauge potential $C^{(3)}\equiv C\omega^{(3)}$
with $\omega^{(3)}$ the CY$_3$ holomorphic volume 3-form.

The kinetic terms of vector multiplets form a special K{\"a}hler manifold
characterized by an analytic prepotential of the ${\cal N}=2$ special
geometry.
Since the dilaton belongs to a hypermultiplet, the prepotential is
determined exactly at the string tree-level. On the other hand, the
kinetic terms of hypermultiplets form a quaternionic manifold, where
radiative corrections are highly restricted by the structure of the
universal hypermultiplet that contains the string coupling. Type IIB is
invariant under S-duality, while in the strong coupling limit of type
IIA, the hypermultiplet space is lifted to 5 dimensions, describing the
complex structure moduli of M-theory compactified on the same Calabi-Yau,
with the dilaton replaced by the CY$_3$ volume.

The one-loop corrections to the hypermultiplet metric were computed
in~\refs{\AntoniadisEG} for directions orthogonal to the dilaton and were
shown to be topological and proportional to the Euler number of the CY$_3$
manifold.  Moreover, they can be easily understood as descending from the
$R^4$ terms in ten dimensions.

In this work, we focus on just the universal hypermultiplet and study the
perturbative string corrections. For this purpose, we can think of type IIA
compactified on a CY$_3$ with no complex structure moduli
($h_{(1,2)}=0$), so that the quaternionic manifold contains only the
dilaton multiplet. The corresponding metric is then reduced to a 4d
self-dual Einstein space of non-zero scalar curvature, while at
tree-level it is further reduced to a symmetric coset space
$SU(1,2)/U(2)$, which is also K{\"a}hler. At a generic order of
perturbation theory, on the other hand, there are only three isometries
corresponding to the three independent shifts of the NS-NS axion and the
complex RR scalar, and they generate a Heisenberg algebra. Imposing just these
isometries, one finds that there is one possible perturbative
correction at the one-loop level, which destroys the K{\"a}hler structure
of the manifold.

Comparing the above supergravity result with the general form of the
effective action in the string frame, including a possible one-loop
correction, one finds an apparent inconsistency with string perturbation
theory.  Moreover, the inconsistency persists even in the absence of
one-loop corrections to the scalar kinetic terms, due to the
one-loop modification of the effective Planck mass that has been
established by a string computation~\refs{\AntoniadisEG}. To resolve this
puzzle, we need to introduce a renormalization of the CY$_3$ volume which
may, in principle, receive contributions even from higher loops. This
phenomenon is analogous to the renormalization of the 4d dilaton in the
presence of higher order $\sigma$-model corrections to the prepotential.

Allowing for such one-loop redefinition of the volume, we find that the
one-loop string effective action can be made compatible with the
quaternionic structure of the universal hypermultiplet for two possible
values of the one-loop correction to the hypermultiplet metric in the
dilaton direction: either zero, or a precise non-zero value proportional
to the Euler number. In the former case, the one-loop correction to the
gravity kinetic terms is absorbed by a shift of the (inverse) string
coupling. However, we show that an explicit one-loop string computation
of the 3-point amplitude involving two RR scalars and one graviton or one
NS-NS antisymmetric tensor selects  the second non-vanishing value
allowed by the field theory analysis. This result makes therefore
the completion of the one-loop correction to the hypermultiplet metric
along the dilaton direction. Moreover, our analysis suggests that the
absence of higher loop perturbative corrections to the hypermultiplet
metric should persist in the presence of additional hypermultiplets
parameterizing the complex structure of the CY$_3$.

Our paper consists of two parts.  In Sections~2 to~7, the physical
implications of the string loop corrections to the universal
hypermultiplet metric are discussed, while the string computation is
contained in Section~5. The second part, composed by three Appendices,
contains the details of the computations needed in the main text. More
precisely, in Section 2, we summarize the generic form of the
four-dimensional type IIA action involving the vectors and
hypermultiplets orthogonal to the volume and dilaton, respectively, as
determined by the analysis of~\refs{\AntoniadisEG}. In Section 3, we
describe the results of Calderbank and Pedersen for the general metric of
one quaternion with two commuting isometries and reduce its form for the
case of three isometries. In Section 4, we compare this metric with the
general form of the one-loop corrected string effective action and show
their compatibility upon introducing a redefinition of the CY$_3$ volume.
In Section 5, we perform a one-loop string computation and determine the
coefficients of the effective action. In Section 6, we discuss the strong
coupling limit which lifts the hypermultiplet space to that of  M-theory
compactified on CY$_3$. In Section 7, we show how our results can be
obtained by reduction of the supersymmetric completion of the $R^4$
couplings in ten dimensions.  Appendix A contains our conventions and
useful properties of the Riemann tensor. We also present the details of
the reduction of the $R^4$ terms from ten to four dimensions.  Appendix B
contains the information about the quaternionic geometry of the universal
hypermultiplet sigma-model and the implementation of string loop
corrections.  Finally, in Appendix C, we present technical details for
the one-loop string computation of the 3-point physical amplitude from
which the one-loop correction to the universal hypermultiplet metric is
extracted.

\newsec{One-loop corrections to the non-universal directions}

In this section, we briefly review the analysis of \refs{\AntoniadisEG}.
We consider type IIA compactified on a CY 3-fold with Betti numbers 
$h_{(1,1)}$ and $h_{(1,2)}$, leading to $h_{(1,1)}$ ${\cal N}=2$
vector multiplets and $h_{(1,2)}+1$ hypermultiplets in four dimensions.  
The one-loop
corrected string effective action in the string frame contains the terms 
\eqn\eSIIa{\eqalign{
S={1\over 2\kappa_4^2}& \int d^4x \sqrt{g^\sigma} \left[\left((1+{\chi_T
\over v_6})\, e^{-2\phi_4} -
\chi_1\right) {\cal R}_{(4)} + \left((1-{\chi_T \over v_6})\, e^{-2\phi_4}
-
\chi_1\right) G_{vv} (\partial v)^2\right.\cr
&\qquad +\left.\left(
(1+{\chi_T \over v_6})\, e^{-2\phi_4} + \chi_1\right) G_{hh}
(\partial h)^2  \right].\cr
}}
$G_{vv}$ is the metric of the $h_{(1,1)}-1$ vector-multiplets orthogonal
to the volume modulus of the internal manifold and $G_{hh}$
is the metric of the $h_{(1,2)}$ non-universal hypermultiplets; $v_6=V_6
(2\pi l_s)^{-6}$ is the normalized volume of the internal Calabi-Yau 
manifold  with Euler number $\chi$ (in string units of length
$l_s=\sqrt{\alpha'})$.  
$\chi_T=2\zeta(3)\chi/(2\pi)^3$ and $\chi_1=4\zeta(2)\chi/(2\pi)^3$ 
are the tree-level and one-loop corrections, respectively\foot{The
normalizations of the $R^4$-action follow those of~\refs{\GreenAS},
where $2\kappa_{10}^2=(2\pi)^7\, l_s^8$, $2\kappa_4^2=2\pi\, l_s^2$. We
refer to the Appendix~A for more detailed discussion of the $R^4$
terms.}.  They descend from the ${\alpha'}^3\, 
R^4$-terms in ten dimensions~\refs{\AntoniadisEG,\GreenAS,\PeetersVW}. 
Here we have dropped world-sheet instanton corrections that are exponentially
suppressed in the large volume limit. The four-dimensional dilaton
$\phi_4$ is related to the ten-dimensional dilaton $\phi_{10}$ via
$e^{-2\phi_4} = v_6\, e^{-2\phi_{10}}$. From now on, we set
$2\kappa_4^2=1$.

For type~IIA reduction on a Calabi-Yau manifold, the universal hypermultiplet
contains the dilaton $\tilde\phi_4$, while the universal vector-multiplet
contains the volume $\tilde v_6$ .  Because there is one conformal compensator
associated with each of these multiplets~\refs{\BerkovitsCB}, $\tilde v_6$ is
the ($\sigma$-model) ``loop" counting parameter for the corrections to the
metric of vector multiplets, and $e^{\tilde\phi_4}$ the ``loop" counting
parameter for the corrections to the metric of hypermultiplets. In ${\cal N}=2$
supergravity, the moduli space 
factorizes into a product of a special K{\"a}hler manifold (vector) 
and a quaternionic manifold (hyper). 
This imposes that $\tilde\phi_4$ and $\tilde v_6$ are a
mixture of the dilaton $\phi_4$ and the volume $v_6$. From general
arguments, it can be easily seen that the tree-level corrections force to
redefine the dilaton, while the loop corrections lead to a
redefinition of the volume
\eqn\eMixing{
e^{-2\tilde\phi_4} \simeq e^{-2\phi_4} \, \left(1+\mu   _T\, {\chi_T\over
v_6}+\cdots\right)~;\quad  \tilde v_6 \simeq v_6 \, \left(1-
{3\mu    _1\over2} \chi_1\, e^{2\phi_4}+ {\cal O}(e^{4\phi_4})\right)
}
No information regarding the mixing can be deduced from~\refs{\AntoniadisEG} since the analysis there was restricted to the
non-universal directions. The value of $\mu    _1$ is determined in Section 4.
An extension of this analysis to tree-level $\zeta(3)\, {\alpha'}^3
\,R^4$-induced corrections to the universal vector-multiplet has been
attempted in \refs{\BeckerNN}, where the tree-level mixing
between the dilaton and the volume was discussed. Unfortunately the
analysis of ref.\refs{\BeckerNN}\
does not allow to derive the precise value of $\mu    _T$. In the
following, we study the metric of the universal hypermultiplet obtained by
compactification of type~IIA/M-theory on CY$_3$, taking into account
one-loop corrections.  This question was partly addressed by
Strominger in~\refs{\StromingerEB}, but with results different from ours.

\newsec{The universal hypermultiplet metric}

\subsec{Calderbank-Pedersen metric}

We now consider the simplest case of just the dilaton hypermultiplet.
Calderbank and Pedersen~\refs{\rfCP} have shown that any self-dual
Einstein metric of non-zero scalar curvature with two commuting
isometries can be derived from a real potential $F(\eta,\rho)$ of two
variables, and has the local form
\eqn\eCP{ ds^2_{CP} = {1\over F^2(\eta,\rho)} \, \left[ \det Q
{d\rho^2+d\eta^2\over \rho^2}  + {1\over \det Q} \pmatrix{d\phi &
d\psi\cr} N^tQ^2 N\pmatrix{d\phi\cr d\psi}\right]
}
with
\eqn\eQN{
Q=\pmatrix{{1\over 2} F - \rho \partial_\rho F& -\rho\partial_\eta F\cr
-\rho \partial_\eta F & {1\over 2} F + \rho \partial_\rho F\cr}~;\qquad
N=\pmatrix{\sqrt\rho&0\cr {\eta\over\sqrt\rho} & {1\over \sqrt\rho}\cr}\,
. }
The metric is Einstein $R_{mn}=3 g_{mn}$ and has a self-dual Weyl tensor 
$W_{rstu}^-=0$  only if the potential
$F(\eta,\rho)$ satisfies the Laplace equation, 
\eqn\eConstraint{
\rho^2 \, (\partial^2_\rho+ \partial^2_\eta) F(\eta,\rho) = {3\over 4} \, F(\eta,\rho).
}
This metric is quaternionic K{\"a}hler, and thus invariant
under local ${\cal N}=2$ supersymmetry. For $\det(Q)<0$, the metric $g^{CP}_{mn}$ is
negative definite with positive scalar curvature, therefore $-ds^2_{CP}$ is
positive definite with a negative curvature scalar ${\cal R}(-g^{CP})=-12$. 
The coupling to the supergravity multiplet 
is~\refs{\ZachosIW,\BaggerTT,\BaggerJF,\DeJaegherKA}  $S= \int
d^4x \sqrt{g^E}\,[R_E - ds^2_{CP}+ {\rm fermions}]$. 

The potential $F(\eta,\rho)$ completely specifies the metric. Its explicit 
form will reflect the loop and D2/M2 instanton corrections, which break
the perturbative shift symmetry of $\eta$. As long as these corrections
are compatible with the constraint~\eConstraint\ the metric remains
quaternionic K{\"a}hler.

\noindent Coupling the quaternionic K{\"a}hler metric~\eCP\ to
gravity and rescaling the space-time metric as $g_{mn} = \sqrt \rho \,
F(\eta,\rho)\,\bar g_{mn}$ gives the effective action
\eqn\eEff{
\int d^4x \sqrt{\bar g}\, \sqrt\rho \, F(\eta,\rho)\, 
\left[\bar {\cal R}_{(4)}- ds^2_{CP}+ {3\over2} \left(\partial_\mu     \ln(\sqrt\rho \,
F(\eta,\rho))\right)^2\right]\ .
}
The tree-level universal hypermultiplet metric
($\chi_1=0$ in~\eSIIa) matches with the quaternionic K{\"a}hler metric~\eCP\ for
$\sqrt{\rho}\, F(\eta,\rho)=\rho^2$ under the field identifications~\refs{\FerraraFF}
\eqn\eClassical{
\rho^2 = e^{-2\phi_4}, \quad C=C_1+i C_2={1\over2}\phi+i\eta, 
\quad \psi=D - C_1C_2\,.
}
$D$ is the scalar obtained after dualization of the NS $B$-field in
four dimensions or the 3-form gauge potential of M-theory in five
dimensions. The classical metric is K{\"a}hler and can be derived from
the  K{\"a}hler potential ${\cal K}=\ln (S+\bar S - 2C \bar C)$ with $S=
\exp(-2\phi_4) + 2i D +C\bar C$. Moreover, it describes a symmetric coset
space with eight isometries $SU(1,2)/U(2)$. For any function $F(\eta,\rho)$
independent of $\eta$, the metric has three U(1) isometries
($\alpha_i\in{\IR}$)
\eqn\eUone{
\phi  \to \phi +\alpha_1~;\quad 
\eta  \to \eta + \alpha_2~;\quad 
\psi  \to \psi + \alpha_3 - \alpha_2 \phi\ .
}
These isometries can be identified with the perturbative Peccei-Quinn
shift symmetries on the Ramond fields $C_{1,2}$ and the NS $B$-field:
\eqn\isometries{
C\to C+{1\over2}\alpha_1+i\alpha_2~;\quad
D\to D+\Bigl({1\over2}\alpha_1 C_2-\alpha_2 C_1\Bigr)
+\Bigl({1\over2}\alpha_1\alpha_2+\alpha_3\Bigr)\, .}
It is easy to verify that the three isometries satisfy the Heisenberg
algebra, $[T_{1,2},T_3]=0$ and $[T_1,T_2]=T_3$, with $T_i$ the generators
associated to the transformation parameters $\alpha_i$ for $i=1,2,3$.

As already mentioned, the quantum corrections to the metric are encoded
in the  solutions to \eConstraint. Here we will be interested only in the
perturbative  corrections.
They must be such that they preserve all three PQ symmetries. 
The only possible deformation of the potential
$F(\eta,\rho)$ compatible with the three U(1) isometries and the
constraint~\eConstraint\ is $\sqrt\rho\,
F(\eta,\rho)=\rho^2-\hat\chi$. 
The string frame expression~\eEff\ suggests the identification
$\sqrt\rho\, F(\eta,\rho)=e^{-2\tilde\phi_4} - \chi_1$, 
with the Planck mass one-loop
correction in~\eSIIa\ expressed in terms of the modified dilaton~\eMixing. 
In ~\refs{\StromingerEB}, Strominger has proposed that all loop corrections
may be captured by modifying the map $\rho^2=f(\exp(-2\tilde\phi_4))$,
while keeping the potential $\sqrt{\rho}\, F(\eta,\rho)= \rho^2$. We will
see below that this is not the case and $\hat\chi$ does not vanish,
implying also the absence of a K{\"a}hler potential. Thus, the one-loop
correction cannot be absorbed by field redefinitions and plays an
important role.

Note that for a potential depending on $\eta$, only two
U(1) isometries ($\alpha_2,\alpha_3$)
are left, and a particular solution to the
constraint~\eConstraint\ is the D-instanton 
function~\refs{\GreenTV,\GreenAS,\PiolineMN,\KehagiasCQ,\GreenBY,\KetovVR}
$F(\eta,\rho)=E_{3/2}$.  Additional wrapped brane instantons (D4 and NS5)
are expected to break all isometries, thus breaking the classical
$SU(2,1)/U(2)$ ~\refs{\FerraraFF,\CecottiQN,\CecottiAD,\BodnerCG,\FerraraIK} to a
discrete set  and should transform this potential into a quaternionic
function F(Q)~\refs{\CecottiAD,\deWitNM,\KasteMV}.

\subsec{Perturbations}

Since we are interested in one-loop corrections to the hypermultiplet 
metric, we will ignore the tree-level corrections proportional to 
$\chi_T=2\zeta(3)\chi/(2\pi)^3$. They only modify the ${\cal N}=2$
prepotential. Therefore the dilaton is not redefined and from now on we
can take $\mu    _T=0$ and $\tilde\phi_4=\phi_4$.

The Calderbank-Pedersen metric for the one parameter family of potentials
preserving the three U(1) isometries,
\eqn\eFI{
F(\eta,\rho) = \rho^{3/2} - \hat\chi\, \rho^{-1/2\,,}
}
has the following form in quaternionic notation: $ds^2_{CP}= -2(u\bar
u+v\bar v)$, with\foot{This is a particular case of the general
expression, valid for $\det(Q)<0$, 
$$
\eqalign{
\sqrt2\, u= {Q_{12}\, d\rho + A\, d\eta\over 2 \rho \, F} 
+i{Q_{11} \alpha + Q_{12}\beta\over F \sqrt{-\det Q}} ~;& \quad
\sqrt2\, v= {A\, d\rho -Q_{12}\, d\eta\over 2 \rho\, F} 
+i{Q_{12} \alpha + Q_{22} \beta\over F \sqrt{-\det Q}} \cr
A=\sqrt{(\rho\, \partial_\rho F)^2 - F^2/4}~;&\quad \alpha=\sqrt\rho\, d\phi~;
\quad \beta=(d\psi+\eta d\phi)/\sqrt\rho\,.}
$$
It is interesting to notice that once instantons are switched on,
$Q_{12}\neq 0$ and the $u$ and
$v$ one-forms will depend both on the RR and NS fields.}
\eqn\eQ{\eqalign{
u&=\sqrt{\rho^2+\hat\chi\over(\rho^2-\hat\chi)^2}\, dC~;\qquad
v=\sqrt{\rho^2\over4(\rho^2+\hat\chi)(\rho^2-\hat\chi)^2}\, (dS+2Cd\bar C)\cr
C&=i\eta+{1\over2}\phi~;\qquad
S=\rho^2+2\hat\chi\ln(\rho)+i\left(2\psi+\eta\phi\right) -C\bar C\,.}}
To compare with Strominger's analysis (Section~6
of~\refs{\StromingerEB}),  we study perturbations of this metric.
We expand the $\rho$ coordinate in a power series of $\exp(2\phi_4)$ as
$\rho^2 =\exp(-2\phi_4) -2\hat\alpha \, \hat\chi + \cdots$.
At the first order, the metric~\eCP\ is modified as 
(using $v+\bar v=2\, d\ln \rho$ at tree level)
\eqn\eSc{\eqalign{
\delta_1 u &={1\over2} (2\hat\alpha+3)\, \hat\chi\, e^{2\phi_4}\ u~;\quad
\delta_1 v =  (2\hat\alpha+1)\, \hat\chi\, e^{2\phi_4}\ v +{1\over2}\, \hat\chi\, e^{2\phi_4}\
\bar v\ .\cr
}}
The $\bar v$ contribution in $\delta_1 v$ was ruled out in
\refs{\StromingerEB}~using the argument that such
contributions introduce parity violating terms (under $D\to
-D,\,C\leftrightarrow\bar C$). This is, however, not the case as long as
$vv$ and $\bar v\bar v$ appear in the metric with equal coefficients.
Strominger's solution corresponds to the potential $F(\eta,\rho)=\rho^{3/2}$.
In Appendix~B, we show that~\eSc\ is a particular case of the most
general physically acceptable one-loop deformation of the metric 
compatible with the quaternionic geometry. 
We also show there that the physically acceptable two-loop deformations
of the metric are captured by the same potential~\eFI, but the map between
$\rho$ and the dilaton has to be modified as
\eqn\eRP{
\rho^2=e^{-2\phi_4}- 2\hat\alpha \, \hat\chi
+\left({\half}-2\hat\gamma+4\hat\alpha\right)\,\hat\chi^2 \, e^{2\phi_4}\ ,
}
with $\hat\gamma$ the only new parameter appearing at two loops (see
Appendix B).
\newsec{Type~IIA string on CY$_3$}

Having established the most general form of the 
perturbative quantum metric compatible with ${\cal N}=2$ supersymmetry, 
we now find the precise identification of the coordinates 
on the universal hypermultiplet target space  with the string variables. 
To do so, we first have to recast \eEff\ in a form that allows a direct 
comparison with a string action (in Einstein frame). We start from the  
metric associated with the one-parameter family of potentials~\eFI:
\eqn\eCPOl{\eqalign{
-{1\over2} \, ds^2_{CP} &={ \rho^2+\hat\chi\over  (\rho^2-\hat\chi)^2} \,
\left((d\rho)^2 +  (d\eta)^2+{(d\phi)^2\over4} \right)+    {\rho^2\over
(\rho^2-\hat\chi)^2 (\rho^2+\hat\chi)}\, (d\psi+\eta d\phi)^2
\ .
}}
As at the tree-level, we introduce the fields
$C={1\over2}\phi+i\eta$ and keeping the definition for $D=\psi +
\eta\phi/2$, we dualize the $dD$ field-strength into a
three-form $H$ by adding the Lagrange multiplier $-{1\over3}\,\int
\epsilon^{\mu    \nu\rho\sigma} H_{\mu    \nu\rho}
\partial_\sigma D$. We thus obtain the following action\foot{We use the
convention for $p$-forms that
$\sqrt{g}\, |F_{(p)}|^2= F_{(p)}\wed * F_{(p)}={\sqrt{g}\over p!}\, F_{\mu   _1\cdots\mu   _p} F^{\mu   _1\cdots\mu   _p}$.}
\eqn\eDualCPi{\eqalign{
S=-2\int d^4x\sqrt{g^E} \, 
&\left[{ \rho^2+\hat\chi\over  (\rho^2-\hat\chi)^2} \, (d\rho)^2
+  {\rho^2 +\hat\chi \over   (\rho^2 -\hat\chi)^2} \, |dC|^2
-  {(\rho^4-\hat\chi^2)(\rho^2-\hat\chi)\over 4\rho^2 }\, |H|^2\right.\cr
&\left. +{i\over2}\epsilon^{\mu    \nu\rho\sigma}
B_{\mu    \nu}\partial_\rho C\partial_\sigma\bar C\right]\cr
}}
We now rescale the RR-scalars\foot{The complexification of RR scalars is a
consequence of string perturbation. The ${\cal N}=(2,2)$ 
$U(1)\times    U(1)$-charge
conservation forbids $dC\, dC$ or $d\bar C \, d\bar C$ terms at one-loop 
so that the
only additional freedom is to rescale the fields $C_{1,2}\to \kappa_{1,2}\,
C_{1,2}$ with $\kappa_1\,\kappa_2 =1$.}
$C_{1,2}$ as $C= f^{-1/2}(\rho) \, C'$ and define $F'= f^{1/2}(\rho)\,
dC$, where $f(\rho)$ is for the moment an arbitrary function. $F'$
satisfies the modified Bianchi identity
$d(f^{-1/2}(\rho)\, F')=0$ solved by $F' =dC' + d\ln(f^{-1/2})\, C'$.
Similarly, we redefine the
$B$-field as $B= f(\rho)\, B'$ and introduce $H'=dB' + d\ln f(\rho)\wed
B'$, so that the last term in~\eDualCPi~remains invariant. This leads to
the action 
\eqn\eCPString{\eqalign{
S = -2 \, \int d^4x \sqrt{g^E}\,&
\left[{ \rho^2+\hat\chi\over  (\rho^2-\hat\chi)^2}\right. \, (d\rho)^2
+ f^{-1}(\rho) {\rho^2 +\hat\chi \over   (\rho^2 -\hat\chi)^2} \, |F'|^2\cr
&\left. - f^2(\rho)\, {(\rho^4-\hat\chi^2)(\rho^2-\hat\chi)\over 4\rho^2 }\,
|H'|^2+{i\over2}\epsilon^{\mu    \nu\rho\sigma}
B'_{\mu    \nu}F'_\rho\,\bar F'_\sigma\right]\ ,
}}
which is to be compared with the string effective field theory.

On the other hand, we consider in the string frame the one-loop corrected
action:
\eqn\eString{\eqalign{
S_1= \int d^4x& \sqrt{g^\sigma} \, \left\{(e^{-2\phi_4}-\chi_1)
R_\sigma + 4\,  \left(e^{-2\phi_4}+\alpha\right)\, (d {\phi_4})^2\right.\cr
&\left.+{1\over2} (e^{-2\phi_4}+\beta) |\hat H|^2-{1\over2}(e^{-2\phi_4} 
+\gamma )\, |\hat F|^2
+{i\over4}\,(e^{-2\phi_4}+\delta)\,\epsilon^{\mu    \nu\rho\sigma} 
\hat B_{\mu    \nu}\,\hat F_\rho\,\hat{\bar F}_\sigma \right\}
}}
where all the coefficients $\alpha, \beta, \gamma$ and $\delta$ are
proportional to the one-loop constant
$\chi_1=4\zeta(2)\chi/(2\pi)^3$ and should be fixed by comparison with
\eCPString. Here we have introduced hatted variables, $\hat B$ and $\hat
C$, which are string variables that have to be related to their
supergravity primed counterparts $B'$ and $C'$ in~\eCPString.

Because of the mixing \eMixing\ between the volume and the
four-dimensional dilaton, the volume dependent terms are also needed:
\eqn\eSv{
S_2 = \int d^4x \sqrt{g^\sigma} \, \left\{-{1\over 6}\,
(e^{-2\phi_4} -\chi_1) \, (d\ln v_6)^2 + \mu    _1 \chi_1\, d\ln v_6 d \phi_4 \right\}
}
This mixing is necessary for the factorization of the vector and
hypermultiplet moduli spaces, as required by ${\cal N}=2$ supergravity. The
factorization also requires the coefficient in front of $(d\ln v_6)^2$ to
be independent of $\phi_4$ in the Einstein frame\foot{ In Appendix~A the
relation between the coefficient $\mu    _1$ and the  $R^4$-terms in ten
dimensions is discussed.}. By redefining the volume vector-modulus as
\eqn\eVt{
\tilde v_6 = v_6 \, \left( 1 - \chi_1 \, e^{2\phi_4}\right)^{3\mu    _1 \over 2}
}
we find that in the string frame the kinetic terms for the volume and the
dilaton take the form:
\eqn\eKDV{
\int d^4x \sqrt{g^\sigma}  \left[4 \left(e^{-2\phi_4}+\alpha +
{3\mu    _1 ^2\over8} {\chi_1^2\over e^{-2\phi_4}-\chi_1} \right)  (d\phi_4)^2  
-{1\over6} (e^{-2\phi_4}-\chi_1) (d\ln\tilde v_6)^2 \right]
}

Finally, in the Einstein frame, the action reads
\eqn\eEinstein{\eqalign{
&S_1+S_2= \int d^4x \sqrt{g^E} \, \left\{ R_E- {1\over 6}\,
\left(d\ln \tilde v_6 \right)^2\right.\cr
&-2 { e^{-4\phi_4}+2(\chi_1-\alpha)\, e^{-2\phi_4} +2\alpha\chi_1-3\mu    _1^2
\chi_1^2/4 \over  (e^{-2\phi_4}-\chi_1)^2} \,
\left(d\phi_4\right)^2+{1\over2}(e^{-2\phi_4}+\beta)(e^{-2\phi_4}-\chi_1)\,
|\hat H|^2 \cr
&\left. -\,{  e^{-2\phi_4}+\gamma\over 2(e^{-2\phi_4}-\chi_1)}\, |\hat F|^2
+{i \over 4} (e^{-2\phi_4} +\delta)
\epsilon^{\mu    \nu\rho\sigma}\hat B_{\mu    \nu}\hat F_\rho\hat{\bar F}_\sigma
\right\}\ .
}} 
Comparison of ~\eCPString\ with ~\eEinstein\ leads to the following
identifications of hatted and primed variables: $(\exp(-2\phi_4)
+\delta)^{1/2}\, \hat F= 2F'$,
$\hat B = B'$ (and $\hat H = H'$). 
Moreover, the matching between the two
metrics appearing in the supergravity quaternionic sigma-model and in
the  Einstein frame 1-loop string effective action gives only two
possible solutions, depending on whether $\hat\chi=0$ or $\hat\chi\neq 0$. 

When $\hat\chi=0$, we recover the solution of \StromingerEB. The field
identifications are the following:
\eqn\eSolutionI{\eqalign{
  \rho^2 &=e^{-2\phi_4}-\chi_1 ,\quad C={1\over2}\phi+i\eta, \quad
\psi=D - C_1C_2,\cr
f(\rho)&=1,\quad
\hat\chi=0,\quad \gamma=\delta, \quad \beta=-\chi_1, \quad
\alpha=\chi_1,\quad \mu    _1 ^2={8\over3}\cr
}}
and the kinetic terms for the modified volume and the dilaton in the
string frame are:
\eqn\eKDVStr{
\int d^4x \sqrt{g^\sigma}  \left[4\, {e^{-4\phi_4}\over e^{-2\phi_4}-\chi_1}\,  (d\phi_4)^2  
-{1\over6} (e^{-2\phi_4}-\chi_1) (d\ln\tilde v_6)^2 \right]\ .
}
The first term was already found in~\refs{\GuntherSC}, but the mixing of the dilaton 
with the volume, necessary for the correct perturbative string interpretation, was not discussed there.

The solution with $\hat\chi\neq0$ is new, and the field identifications are 
\eqn\eSolutionII{\eqalign{
  \rho^2 &=e^{-2\phi_4}-\chi_1 ,\quad C={1\over2}\phi+i\eta, \quad
\psi=D - C_1C_2,\cr
f(\rho)&=1-\chi_1\, e^{2\phi_4},\quad
\hat\chi=-\chi_1,\quad \gamma=\beta=-2\chi_1, \quad
2\alpha=5\chi_1,\quad\delta=0,\quad \mu    _1 ^2=4\cr
}}
Both solutions are consistent with ${\cal N}=2$ supergravity, although
only one solution corresponds to the low energy effective action of string
theory. The one-loop corrections to the kinetic term of the RR fields and
to the Chern-Simons term, depend on the parameters of the effective action
\eString.  In the context of quantum
field theory $\gamma$ and $\beta$ are wave-function renormalization, and
$\delta$ the vertex correction. Field-theoretically the
wave-function renormalization is fixed by the two-point function, but the
closed string two-point amplitude vanishes on-shell and does not have a
reliable off-shell extension. Therefore, we have to consider the three-point
amplitude which computes the $S$-matrix elements corresponding to the
renormalized couplings: $\langle GF\bar F\rangle=\chi_1/2$ and
$\langle BF\bar F\rangle=\delta-\gamma-\beta/2$. 
For the solution~\eSolutionI\ the tree-level and one-loop $S$-matrix elements are the same.
We show by a direct one-loop string computation presented in Section~5
that  this is not the case. This selects the solution~\eSolutionII. Thus,
one concludes that the one-loop corrections to the universal
hypermultiplet metric are physical. We will discuss the higher-loop
corrections in Section~6. Note that the solution has a sign ambiguity
$\mu _1 = \pm 2$, which we cannot determine by our analysis.

\medskip
\noindent
{\it Remarks about the solution~\eSolutionII}
\smallskip

\item{$\triangleright$} The identification of the supergravity and
string metrics required field redefinitions for both the NS-NS and RR
fields: 
\eqn\eBI{
\hat H=d\hat B-2{\chi_1 \over e^{-2 \phi_4} - \chi_1} d\phi_4 \wed\hat B\quad;
\quad\hat F=2e^{\phi_4}\, (1 -\chi_1 \, e^{2\phi_4})^{1/2} \, dC\ . }
These identifications in turn imply modifications of the Bianchi 
identities by terms proportional to $\chi_1$.
In the background of the Calabi-Yau manifold, 
there is a non-trivial dilaton, 
$\phi_{10}=\phi_{10}^{o} - e^{2\phi_{10}^{o}} \, 2\zeta(2)\, E_6/ (2\pi)^3$,
which leads to
\eqn\eDD{
e^{-2\phi_4}=e^{-2\phi_{4}^{o}}\, \left(1+\chi_1\, e^{2\phi_{4}^{o}}\right) \ .
}
Therefore the field redefinition of the RR field strengths
\eqn\eRR{
\hat F=2\left(e^{2\phi_4}\, (1-\chi_1\, 
e^{2\phi_4})\right)^{1/2}\, dC=e^{\phi_{4}^{o}}\, dC
}
is the standard redefinition for this particular dilaton background. As
for the modified Bianchi identity for the NS field, it simply leads to
new interaction terms between the $B$-field and the dilaton, needed for
supersymmetry of the one-loop action.

\item{$\triangleright$} There is one more crucial distinction between 
\eSolutionI\ and \eSolutionII.  By an interesting coincidence, the
classical metric on $SU(2,1)/U(2)$ happened to be K{\"a}hler. It can be
easily checked  that the metric~\eCPOl\ admits a  closed K{\"a}hler
two-form only if $\hat\chi=0$. Therefore the K{\"a}hler character of the
tree-level metric (defined with $F(\eta,\rho)=\rho^{3/2}$) is lost once
quantum corrections are turned on.

\newsec{Reconstructing the four-dimensional effective action}

\subsec{Three-point S-matrix elements}

In this Section, we perform a string one-loop computation which 
allows us to distinguish between the two solutions \eSolutionI\ and 
\eSolutionII\ which determine the effective action~\eString.
One possibility is to study the kinetic terms of the NS 2-form field $B$.
The one-loop correction to the $H^2$-metric arises from the one-loop term
${\alpha'}^3\, R^3 H^2$ obtained in~\refs{\PeetersUB} (given in the
string frame):
\eqn\eCa{\eqalign{
\int d^{10}x \sqrt{G}\, \delta^{r_1\cdots r_9}_{s_1\cdots s_9}\,
R^{r_1r_2}_{s_1s_2} R^{r_3r_4}_{s_3s_4} R^{r_5r_6}_{s_5s_6}&
\left(H_{r_7r_8s_9} H_{s_7s_8r_9} -   {1\over 9} H_{r_7r_8r_9} H_{s_7s_8s_9}
\right)\cr
&\to \chi \, \int d^4x \sqrt{g^\sigma}\, H_{r_1r_2r_3} H^{r_1r_2r_3}\ .
}}
This shows that $\beta\propto \int_{M_6}\, R\wed R \wed R\sim \chi$ is
proportional to the Euler characteristic of the Calabi-Yau manifold.
Unfortunately, because $S$-matrix elements evaluate the renormalized
couplings, the three-point amplitude $\langle BBh\rangle= \beta - 2
\times \beta/2 + \chi_1/2$ (see below) does not contain any information
about the value of $\beta$. The precise relation can be fixed by the
analysis of a four point amplitude in four dimensions, but we will not do
this here.  We can nevertheless find some other appropriate three-point
amplitudes which will make possible to discriminate between 
\eSolutionI\ and \eSolutionII. 

The field theory limit of the three-point one-loop amplitude decomposes
into the three diagrams given in fig.~1, where the blob corresponds to the
one-loop correction
coefficients\foot{See~\refs{\PeetersVW,\PeetersUB,\GrossMW,\ForgerVJ}  
for a rational about extracting the low energy string effective action
from $S$-matrix elements.} in the low-energy string effective
action~\eString.  
If  the external lines are RR-scalars and the
wiggly line the graviton, the amplitude is
\eqn\eGFF{
<G F \bar F> = {1\over2}F_\mu    \, h^{\mu    \nu} \, \bar F_\nu\, \left(\gamma - 2 \times 
{\gamma\over2} + {\chi_1\over2} \right)\ .
}
On the other hand, the correction to the Chern-Simons coupling is
\eqn\eBFF{
<B F \bar F> ={i\over4}\, \epsilon^{\mu    \nu\rho\sigma} B_{\mu    \nu}\,F_\rho \,\bar F_\sigma\,
\left(\delta - 2 \times     {\gamma\over2} - {\beta\over2} \right)\ .
}
Thus, computing these string amplitudes will allow us to decide which of
the two solutions for the hypermultiplet metric is chosen by string
theory. 

$$
\vcenter{\vbox{\psfig{file=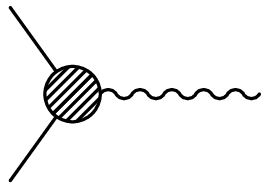}}}+2\times    \,
\vcenter{\vbox{\psfig{file=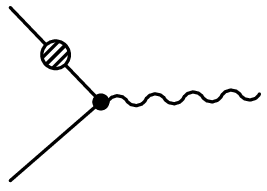}}}+
\vcenter{\vbox{\psfig{file=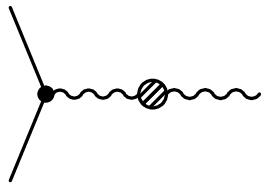}}}
$$
\fig\IIIpoints{}%
{\narrower~\noindent
\ninesl The field theory diagrams contributing to the
three points S-matrix reconstructed from the effective action~\eString.
The first diagram represents the one-loop correction to the vertex
function, the second and the third the wave-function
renormalization.\smallskip}

\subsec{Definition of the vertex operators}

We now compute the same scattering amplitudes in type IIA string theory 
at one loop. This requires the computation of the three-point 
torus correlation function 
of the vertex operators for the RR field strengths $F$, 
$\bar F$ and the NS field (either the graviton or the anti-symmetric
tensor). For the comparison with the field theory result we need
the amplitude to ${\cal O}(k^2)$. Useful references for the one-loop 
calculations are \refs{\AtickRS,\Minahan,\BernPK}. We use the conventions
of \AtickRS\ which means e.g. that $\alpha'=2$.

The NS-NS vertex operator is\foot{Tilded fields are right movers.}
\eqn\eVNS{
V_{\rm NS}^{(-1,-1)}=\zeta_{\mu   \nu}\,:\!\psi^\mu   \tilde\psi^\nu e^{ik\cdot   X}
e^{-(\varphi+\tilde\varphi)}\!:}
where $\zeta_{\mu    \nu}$ is traceless and symmetric for the graviton $h$ and 
antisymmetric for the antisymmetric tensor $B$. 
The vertex operators of the RR-scalars in the type IIA 
universal hypermultiplet in the
$(-{1\over2},-{1\over2})$-ghost picture are composed of  
left- and right-moving $SO(4)_{\rm space-time}$ spin fields $S$ and $\tilde S$ 
and internal ${\cal N}=(2,2)$ superconformal fields $\Sigma$ and its 
conjugate $\bar\Sigma$
\eqn\eVRR{\eqalign{
V^{(-{1\over2},-{1\over2})}_F &
=\hat F_\mu    ~:\!S^\alpha (\sigma^\mu    )_{\alpha\dot\beta}
{\tilde S}^{\dot\beta} \ \Sigma\
e^{-{1\over2}(\varphi+\tilde\varphi)}\, e^{ik\cdot    X}\!: \cr
V^{(-{1\over2},-{1\over2})}_{\bar F} &=\hat{\bar  F}_\mu      ~:
\!{S}_{\dot\alpha}(\bar\sigma^\mu    )^{\dot\alpha\beta}  \tilde S_{\beta}
\ \bar\Sigma e^{-{1\over2}(\varphi+\tilde\varphi)}\, e^{ik\cdot    X}\!:\ .\cr}}

To compute the correlation function, we rotate to Euclidean 
signature and bosonize, as usual, the complex
world-sheet fermions and spin-fields according to 
$\psi^I\sim e^{i\phi^I},\,\psi^{\bar I}\sim e^{-i\phi^I}$ $(I=1,2)$
and $S_{\alpha}\sim e^{\pm    {i\over2}(\phi^1+\phi^2)}$, 
${S}_{\dot\alpha}\sim e^{\pm    {i\over2}(\phi^1-\phi^2)}$. 
For simplicity, in these bosonization formulas we omit cocycle factors
which are needed to obtain $SO(4)$ covariant correlation functions.
Finally, the spinor indices are raised and lowered with the epsilon-tensor
$\epsilon^{\alpha\beta}$ and $\epsilon^{\dot\alpha\dot\beta}$, with the
conventions:
\eqn\eSpinors{\eqalign{
S_{(--)}=e^{-{i\over2} (\phi^1+\phi^2)} = S^{(++)}, &\quad \dot S_{(+-)}=e^{{i\over2} (\phi^1-\phi^2)} = \dot S^{(-+)}\cr 
(\sigma^1)_{--,-+}= (\bar\sigma^1)^{+-,++}&,\quad 
(\sigma^{\bar 1})_{++,+-}= (\bar\sigma^{\bar 1})^{-+,--}\cr
(\sigma^2)_{--,+-}=(\bar\sigma^2)^{-+,++}&,\quad
(\sigma^{\bar 2})_{++,-+}= (\bar\sigma^{\bar2})^{+-,--} \ .
}}

\subsec{The tree-level amplitude}

The  tree-level amplitude is given by\foot{We use Wess and
Bagger~\refs{\WessCP} conventions $\sigma^a \bar \sigma^b + \sigma^b \bar
\sigma^a =-2\eta^{ab}$ and the identities
${\rm tr}(\sigma^a \bar \sigma^b \sigma^c \bar\sigma^d) =
2(\eta^{ab}\eta^{cd}+\eta^{bc}\eta^{ad} -
\eta^{ac}\eta^{bd}-i\epsilon^{abcd})$ and $\sigma^a \bar \sigma^b\sigma^c=\eta^{ac} \sigma^b - \eta^{bc} \sigma^a
-\eta^{ab} \sigma^c+i\epsilon^{abcd} \sigma^d$.}

\eqn\eATree{\eqalign{
\langle V_{NS}^{(-1,-1)}(x) V_F^{(-\half,-\half)}(u) V_{\bar
F}^{(-\half,-\half)}(v)\rangle &\sim {\rm
tr}(\bar\sigma^\kappa\sigma^\nu\bar\sigma^\lambda\sigma^\mu   )\,
\zeta_{\mu   \nu}\, F_\lambda\, \bar F_\kappa\cr
&=2[2 F\cdot   h\cdot   \bar F- i \epsilon^{\mu   \nu\rho\sigma} B_{\mu   \nu} F_\rho
\bar F_\sigma]
}}
where we have used 
\eqn\eClebsch{
\psi^\mu   \, S_\alpha\, S_{\dot \beta}\sim
{1\over\sqrt{2}}{(\sigma^\mu   )_{\alpha\dot\beta}}\, , \qquad \psi^\mu  
\, S^{\dot\alpha}\, S^{\beta}\sim{ 1 \over\sqrt{2}}(\bar\sigma^\mu  
)^{\dot\alpha\beta}}
The one-loop amplitude will be again a linear combination of the
same tensorial structures. The task is to decide whether they come with 
the same relative coefficients as the tree-level or not. This will be the 
criterion to decide between the two solutions we found in Section~4.

\subsec{The one-loop amplitude}

In Appendix C, we give some technical details and useful formulae for the
one-loop computation, and we discuss the independence of the physical
amplitude from the supercurrent insertion points. Here,
we start from the representation with all the vertex operators in the
canonical ghost picture. 

On a toroidal world-sheet, the left- and right-moving superconformal 
ghost charges have to add up to zero separately. If all vertices are 
chosen in the canonical ghost picture --- $(-1,-1)$ for NS-NS fields and 
$(-\half,-\half)$ for RR-fields --- we have also to insert two
left-moving and  two right-moving picture changing operators, $Y(z)$ and 
$\tilde Y(\bar w)$. We thus have to compute\foot{The existence
of one conformal Killing vector on the torus allows to fix the position of 
one of the three vertex operators. We can instead use translational invariance 
to integrate over all three positions and compensate by dividing with
$\tau_2$, the volume of the torus.}
\eqn\eADef{\eqalign{
&{\cal A}=\sum_{s,\tilde s=1}^4
\int_{\cal F}{d^2\tau\over\tau_2}Z(\tau,\bar\tau)
\int\!d^{2}\!x\int\!d^{2}\!u\int\!d^{2}\!v~
\langle Y(z_1) Y(z_2) \tilde Y(\bar w_1) \tilde Y(\bar w_2)\cr
&\qquad\qquad\qquad\qquad 
\times   V_{\rm NS}^{(-1,-1)}(x) V_{F}^{(-{1\over2},-{1\over2})}(u)
V_{\bar F}^{(-{1\over2},-{1\over2})}(v)\rangle_{s,\tilde s}}}
where $Z(\tau,\bar\tau)=(64\pi^4\tau_2^2|\eta(\tau)|^4)^{-1}$ is the
bosonic  partition function, and the summation is over  all possible 
$4\times 4$ spin-structures $s, {\tilde s}$. We have normalized the
short-distance singularity of bosonic correlators to one, while those for
fermions and super-conformal ghosts to their respective partition
functions.

The relevant part of the picture changing operators is 
$Y=e^{\varphi} T_F$ where $T_F=\partial X^\mu   \psi_\mu   +T_F^{\rm int}$ is the
matter part of the world-sheet supercurrent and $\varphi$ and $\bar\varphi$ belong to 
the `bosonized' superconformal ghost system.   
For compactifications on orbifolds  $T_F^{\rm int}=\sum_{I=3,4,5} \partial
X^I\psi^{\bar I}$.  
We have to compute the following correlation function 
\eqn\eACanonical{
\langle e^\varphi T_F(z_1) e^\varphi T_F(z_2) 
e^{\tilde\varphi}\tilde T_F(\bar w_1) e^{\tilde\varphi}\tilde T_F(\bar w_2) \,
V_{NS}^{(-1,-1)}(x) V_{F}^{(-\half,-\half)}(u) V_{\bar F}^{(-\half,-\half)}(v)\rangle_{s,\tilde s}}
We choose the polarizations of the RR and NS-NS states such that the tree-level
amplitude is zero. At the level of conformal field theory this is achieved by
non-conservation of $SO(4)$ charge. None of the two Lorentz structures appearing on the 
r.h.s. of \eATree\ vanishes separately but they cancel\foot{We have written 
\eATree\ in Minkowski signature while the calculation below is performed 
in Euclidean signature.}. 
The criterion which decides between the two solutions
\eSolutionI\ and \eSolutionII\ is whether the one-loop amplitude also vanishes 
for this choice of polarization or not. Before embarking on the one-loop
computation we remark that for this choice of polarizations there is no
contribution from the second diagram
of~\IIIpoints\ since it is proportional to the tree-level vertex~\eATree.
An appropriate choice which leads to a vanishing tree-level amplitude is:
\eqn\eTrickH{
\zeta_{12}\, F_{\bar 1}\, \bar F_{\bar 2}\, 
\langle\psi^1(x) S^L_{(--)}(u) \dot S^L_{(+-)}(v)\rangle_s\,\,
\langle\tilde\psi^2(\bar x) \dot S^R_{(-+)}(\bar u) S^R_{(--)}(\bar v)\rangle_{\tilde s}
} 

The one-loop amplitude then becomes
\eqn\eACanonical{\eqalign{
&{\cal A}_1=\zeta_{12}  F_{\bar 1}\bar F_{\bar 2} \sum_{s,\tilde s} \int_{\cal
F} {d^2\tau\over\tau_2}\, Z(\tau,\bar\tau)\int d^2x\,d^2u d^2v\,
\langle\Sigma(u)\bar\Sigma(v)\rangle_s\langle\tilde{\bar\Sigma}(\bar u)\tilde\Sigma(\bar v)\rangle_{\tilde s}\cr
&\times \langle\partial X^{\mu _1}(z_1)\partial X^{\mu   _2}(z_2)\bar\partial 
X^{\nu_1}(w_1)\bar\partial X^{\nu_2}(w_2) \prod_{i=1}^3 e^{i k^{(i)}\cdot   X}\rangle\cr
&\times \langle \psi^{\mu _1}(z_1)\psi^{\mu _2}(z_2)\psi^{1}(x)~S^L_{(--)}(u)\dot
S^L_{(+-)}(v)\rangle_s\langle e^{\varphi(z_1)}e^{\varphi(z_2)}e^{-\varphi(x)}
e^{-{\varphi(u)\over2}}e^{-{\varphi(v)\over2}}\rangle_s\cr
&\times    \langle \tilde\psi^{\nu_1}(\bar
w_1)\tilde\psi^{\nu_2}(\bar w_2)\tilde\psi^{2}(\bar x) \dot S^R_{(-+)}(\bar
u)S^R_{(--)}(\bar v)\rangle_{\tilde s} \langle e^{\tilde \varphi(\bar
w_1)}e^{\tilde \varphi(\bar w_2)} e^{-\tilde\varphi(\bar x)}e^{-{\tilde\varphi(\bar u)\over2}}
e^{-{\tilde\varphi(\bar v)\over2}}\rangle_{\tilde s}\cr}}
with arbitrary positions of the four picture changers. The only choices
of indices for the fermions which lead to non-vanishing fermionic
correlation  functions are, in a complex basis,  
$(\mu   _1,\mu   _2)=(\bar 1,2)$ or $(2,\bar 1)$ and $(\nu_1,\nu_2)=(\bar 2,1)$ or
$(1,\bar 2)$\foot{In our choice of polarizations for
the external states, the internal part of the supercurrent cannot
contribute.}. 
We consider the choice $(\mu _1,\mu _2,\nu_1,\nu_2)=(\bar 1,2,\bar 2,1)$. 
The others are obtained (up to a sign) by exchanging $z_1\leftrightarrow
z_2$ and/or $w_1\leftrightarrow w_2$.

The fermionic contractions give
\eqn\eC{\eqalign{
C_s^L &\equiv{K\over \sqrt8}\delta_s {\theta_1(z_1-u)\theta_1(z_2-x)\over\theta_1(x-u)\theta_1(z_1-z_2)}\,
{\theta_s(x-z_1-\Half(u-v))\theta_s(z_2-\Half(u+v))\over
\theta_s(z_1+z_2-x-\Half(u+v))\theta_1(u-v)}\, \prod_{I=3}^5 \theta_{s,{\bf h}_I}(\Half(u-v))\cr
C_{\tilde s}^R&\equiv{\bar K\over \sqrt8}\bar\delta_s{\theta^*_1(\bar w_1-\bar v)
\theta^*_1(\bar w_2-\bar x)\over\theta^*_1(\bar x-\bar v)\theta^*_1(\bar w_1-\bar w_2) }\,
{\theta^*_s(\bar x-\bar w_1-\Half(\bar u-\bar v ))\theta^*_s(\bar w_2-\Half(\bar u+\bar v))\over
\theta^*_s(\bar w_1+\bar w_2-\bar x-\Half(\bar u+\bar v))
\theta^*_1(\bar u-\bar v)}\,\prod_{I=3}^5 \theta^*_{s,{\bf h}_I}(\Half(\bar u-\bar v))\ .\cr
}}
The relative phases  $\delta_s$ and $\bar\delta_s$ are determined by requiring
periodicity in all position variables on the world-sheet torus.
The normalization constants are obtained from the short distance behavior of the amplitudes and are found to be 
\eqn\eCf{
K= \langle {\bf 1}_{\psi+\varphi}\rangle_s \, {{\theta_1'}^{3\over2}(0)\over
\theta_s(0)} = {\theta_1'(0)\over \eta(\tau)}=-2\pi \, \eta^2(\tau)=\bar K^*\ .}
Where ${\bf 1}_{\psi+\varphi}$ is the unit operator the space-time fermions
and the superconformal ghosts. Sending $z_1\to x$,  and $w_1\to x$  and  fixing
$\delta_1=1$, we get $\delta_s=(-1)^s$ for $s=2,3,4$ (likewise for the
right-movers). The sum over  spin structures can then be done using the
Riemann identity (see Appendix C) with the result 
\eqn\eSum{
\sum_{s,\tilde s}C^L_s C^R_{\tilde s}=2\pi^2|\eta|^4\,
\prod_{I=3}^5\left|\theta_{1,-{\bf h}_I}(0)\right|^2\ .}
This is related to the Euler characteristic of the compactification
manifold by remarking that $\prod_{I=3}^5 \theta_{1,-{\bf h}_I}(0) = {\rm
Tr}_{R}(-)^{F_L}$ and that ${\rm Tr}_{RR}(-)^{F_L+F_R}= \chi$

In the same limit at least one of the fermionic correlators for 
any of the other three possible index structures vanishes. 
In the choice \eACanonical, the only possible bosonic contractions are
\eqn\eBos{\eqalign{
&\langle\partial X^{ 1}(x)\bar\partial X^{\bar 1}(w_2) \partial X^{\bar 2}(z_2) \bar\partial X^{ 2}(x) \prod_{i=1}^3 \,
e^{ik^{(i)}\cdot   X}\rangle\cr
&=\langle\partial X^{ 1}(x)\bar\partial X^{\bar 1}(w_2)\rangle
\langle\partial X^{\bar 2}(z_2)\bar\partial X^{2}(x)\rangle
\langle\prod_{i=1}^3 e^{k^{(i)}\cdot   X}\rangle\cr
&~-\langle \bar\partial X^{\bar 1}(w_2) \partial X^{\bar 2}(z_2)\prod_{i=1}^3
\, e^{ik^{(i)}\cdot   X}\rangle\,  \left( k^{(2)}_{\bar 1}\partial_x
G_B(x-u) +  k^{(3)}_{\bar 1} \partial_x
G_B(x-v) \right)\cr
&\qquad\qquad\times    \left( k^{(2)}_{\bar 2}\bar\partial_x
G_B(x-u) +  k^{(3)}_{\bar 2} \bar\partial_x G_B(x-v) \right)}}
The first term on the r.h.s. (second line) contributes only through zero
modes, while the remaining terms are of order ${\cal O}(k^4)$. 
Indeed, for generic positions $z_2$ and $w_2$ of the supercurrent
insertions, the integration over the positions of the vertex operators
cannot lead to  cancellation of these momentum factors. We are thus left
with the first term which leads, upon  integration over the positions
$x$, $u$ and $v$ of the vertices, to 
\eqn\eAFinal{\eqalign{
{\cal A}_1&= 2\pi^4\, \chi\,\zeta_{ 1  2} \, F_{\bar 1}\,\bar F_{\bar 2} \,
\left[\int_{\cal F} d^2\tau \, Z(\tau,\bar\tau)\, |\eta|^4+{\cal O}(k^2)\right]\cr
&=  {\pi^2\over 3\cdot   2^5}\,
\chi\,\zeta_{12} \, F_{\bar 1}\,\bar F_{\bar 2}+|F|^2 \,{\cal O}( k^2) \ .
}}
This shows that the one-loop amplitude is non-zero for a configuration that
makes the tree-level amplitude to vanish. We conclude that the
string one-loop correction to the universal hypermultiplet is given by the solution~\eSolutionII.

\newsec{M-theory on CY$_3$}

In this Section we discuss the M-theory lifting, or the strong coupling
limit of the previous construction. Of course, the
expression~\eCP\ should now describe the metric for the universal
hypermultiplet obtained by compactification of $l_P^6\, R^4$-corrected
M-theory to five-dimensions on a Calabi-Yau three-fold. However, the
identification of coordinates in ~\eCP\ with M-theory variables is subtle.

In five dimensions, the universal hypermultiplet is composed
of~\refs{\CadavidBK} the (normalized) volume of the Calabi-Yau 
manifold $\hat v_6=(\int\!\sqrt{g}d^6x)\, l_P^{-6}$, the three-form 
$C_{\mu   \nu\rho}$, which in 5d is dual to a scalar, and the complex scalar $\hat
C$ obtained from the RR 3-form along the volume form of the CY$_3$,
$C^{(3)}={\hat C}\omega^{(3)}$. The four-dimensional universal vector
multiplet (which contained
$\tilde v_6$) upon lifting becomes part of the gravity multiplet.
Reduction of the $l_P^6 R^4$-corrected action on a Calabi-Yau three-fold
to five dimensions gives then the following universal hypermultiplet metric
\eqn\eMth{\eqalign{
&S=\int d^5x \, \sqrt{G}\, \left[(\hat v_6-\chi_1)\, {\cal R}_{(5)}+
{5\over 6}\, (\hat v_6+\hat\alpha ) \, (d\ln\hat v_6)^2- {1\over2} (\hat
v_6 + \hat\beta) (F_4)^2\right. \cr &\left. -{1\over 2} (\hat
v_6+\hat\gamma) |\hat F|^2 +i (\hat v_6 +\hat\delta) C_3 \wed \hat F \wed
\hat{\bar F} \right]\ , }}
where $F_4$ is the field-strength of the 5d RR 3-form $C_3$. Because the
relative normalization between  ${\cal R}_{(11)}$ and $R^4$ terms in eleven
dimension is the same as between the ${\cal R}_{(10)}$ and the one-loop $R^4$ for
type~IIA in ten dimensions~\refs{\GreenAS}, the Planck mass correction is
again given by $\chi_1$.

In order to compare this action with the strong coupling limit of
the low energy string effective action determined in the previous
section, we should first stress their differences. Since in the string
frame supersymmetry remains exact loop by loop, the string action
~$\eString + \eSv$ is a one-loop exact supersymmetric effective action. 
Such a statement cannot be made for M-theory compactifications; rather,
the effective action~\eMth\ should be thought as a large-$\hat v_6$
approximation of an exact supersymmetric action. We saw that in four
dimensions a mixing of the dilaton with the volume, or in other words of
the universal hyper- and vector multiplets,  was necessary for obtaining
a supersymmetric string effective action. As already mentioned, M-theory
does not have a universal vector multiplet and thus the matching between
the action~\eMth\ and the Calderbank-Pedersen metric~\eCP\ can only be
done up to the order
${\cal O}(\hat v_6^{-1})$. 

For the identification of~\eMth\ with the strong coupling limit of~$\eString +
\eSv$ we need  the standard relation~\refs{\WittenEX} between the
five-dimensional (M-theory) and the four-dimensional (string) metrics:
\eqn\eMetric{
R_5^{-1}\,l_s^{-2} \,  ds^2_{(4)} = l_P^{-2} \, [ ds^2_{(5)} - R_5^2\, \,
(dx^5 - C_\mu       \, dx^\mu       )^2]\, .
}
We thus identify the dilaton with the volume of the Calabi-Yau
measured in $l_P$ units
\eqn\eMapI{
 \hat v_6= e^{-2\phi_4}\ ,
}
and the members of the universal hypermultiplets are identified by 
$(\hat v_6, C_{\mu       \nu\rho}, \hat C_1,\hat C_2) \to (\phi_4, R_5^{-1}\,
B_{\mu       \nu}, R_5^{-3/2} \, C_1, R_5^{-3/2} \, C_2)$.
The reduction to four dimensions of the volume dependent part of~\eMth\
then gives
\eqn\eReduc{\eqalign{
S_v =\int d^4x \, \sqrt{g} \, &\left\{(\hat v_6-\chi_1) \left[{\cal
R}_{(4)} -{1\over6} (d\ln[ R_5^3( \hat v_6-\chi_1)])^2\right]\right. \cr
&\left.+ (\hat v_6 + {5\hat\alpha+2\chi_1\over 6} + {1\over 6} {\chi_1^2\over \hat v_6-\chi_1}) (d\ln
\hat v_6)^2\right\}
}}
This matches with the four-dimensional action determined in the previous
section upon the following identification 
\eqn\eMapII{
R_5^3\, (\hat v_6- \chi_1) = \tilde v_6\ ,
}
which implies that the strong coupling radius $R_5$ is
modified by a volume dependent term (in M-theory units) as
\eqn\eR{
R_5^3\simeq e^{2\phi_{10}} \, \left(1-\left({{3\mu    _1  \over2}-1}\right) \,
{\chi_1\over\hat v_6}+{\cal O}(\hat v_6^{-2}) \right)
}
This identification, up to terms of order ${\cal O}(\hat v_6^{-1})$, 
allows us to fix all the constants in~\eMth\ as
$5\hat\alpha+2\chi_1=6\alpha$, $\hat\beta=\beta$, $\hat\gamma=\gamma$ and
$\hat\delta=\delta$, but does not provide any information about $\mu   _1$. 
\newsec{Supersymmetry and higher-loop terms}

As we have seen, the only perturbative corrections to the universal
hypermultiplet are at one loop, and all other contributions from higher
loop/derivative terms can be absorbed in field
redefinitions. Similarly, non-universal hypermultiplets and vector
multiplets get corrected only at one-loop and tree-level respectively
\refs{\AntoniadisEG}.  All these corrections can be seen as descending
from ten-dimensional $R^4$
terms~\refs{\AntoniadisEG,\BeckerNN,\AntoniadisTR} (see Appendix A for
such a derivation for the universal hypermultiplet). Clearly
string theory has many more terms which are higher-order in curvature,
may contain higher numbers of derivatives, and appear at higher loops.
The conclusion of our analysis in Appendix B is that, when reduced to
four dimensions, the contributions from all these extra terms should be
absorbable in field redefinitions of the dilaton. Here, we would like to
examine what constraints are imposed by this property on certain higher
loop/derivative terms in ten dimensions.  Naturally, such indirect
analysis can apply only to a very limited set of couplings.

The only terms that can possibly be constrained by such analysis are
those that upon reduction to four dimensions survive the large volume
limit.  Using $e^{-2\phi_4} = v_6\, e^{-2\phi_{10}}$, it is not hard
to see that such terms are of the type $R^{3m+1}$ (and $H^2R^{3m}$ or
$F_{RR}^2 R^{3m}$) where $m$ counts the loops. In other words, these
are exactly the same terms that lift to eleven dimensions. It is not
very hard to see that at two-derivative level the only contribution
comes from strictly factorized terms of the form ${\cal R} \, {\cal
S}^{(m)},$ where ${\cal R}$ is the Ricci scalar and ${\cal S}^{(m)}
\sim R^{3m}$ is a fully-contracted combination of Riemann tensors.
Similarly, corrections to the $B$-field kinetic term come from
fully-factorized terms in ten dimensions. A detailed discussion of the
$m=1$ case can be found in Appendix A, and ${\cal S}^{(1)}$ coincides
with the six-dimensional Euler density. 
These contributions in the ten-dimensional action have a   priori
ambiguities, due to possible field redefinitions, but are constrained
by the ${\cal N}=2$ supersymmetry in four dimensions. In order for the
higher-order corrections to the universal hypermultiplet metric to be
absorbable from two loops and higher, the corrections to the
four-dimensional Einstein term and the kinetic term for the $B$-field
must be exactly the same and of the form: 
\eqn\tendone{S= \int d^{10}x \sqrt{g^\sigma} ({\cal R}_{(10)} +
{1 \over 2} H \wed *H) \sum_{m \ge 2} (e^{-2\phi_{10}})^{1-m}
{\cal S}^{(m)}} 
The special geometry of ${\cal N}=2$ supersymmetry requires
that the integrals over the internal manifolds $M_6$, 
$\int_{M_6} {\cal S}^{(m)}\equiv \delta^R_m$, are independent of K{\"a}hler moduli.

We can further specify the precise form of ${\cal S}^{(m)}$ in 
\tendone. To this end, we turn to the non-universal sector and
consider the higher-order corrections to \eSIIa:  
\eqn\eSIIaHL{\eqalign{ S=& \int d^4x \sqrt{g^\sigma}
\left[\left((1+{\chi_T \over v_6})\, e^{-2\phi_4} - \chi_1 + \sum_{m \ge 2}
(e^{-2\phi_4})^{1-m} \delta^R_m \right) {\cal R}_{(4)} \right.\cr 
&\qquad +\left. \left((1-{\chi_T \over v_6} )\, e^{-2\phi_4} - \chi_1 + \sum_{m 
\ge 2} (e^{-2\phi_4})^{1-m} \delta^V_m \right) G_{vv} (\partial 
v)^2\right.\cr
&\qquad +\left.\left( (1+{\chi_T \over v_6})\, e^{-2\phi_4} + \chi_1 +
\sum_{m \ge 2} (e^{-2\phi_4})^{1-m} \delta^H_m \right) G_{hh} (\partial
h)^2 \right]\, ,}} 
where as before, $G_{vv}$ is the metric of the vector-multiplets orthogonal
to the volume modulus of the internal manifold and $G_{hh}$ is the metric
of the hypermultiplets orthogonal to the universal direction.  In order to
be able to absorb all higher-loop corrections by redefining the
dilaton, clearly 
$\delta^R_m=\delta^V_m=\delta^H_m$
should hold for all $m \ge 2$. Moreover, in order not to spoil the metric
on the hypermultiplet moduli space, $\delta^R_m = \int_{M_6} {\cal S}^{(m)}$ should
be independent of complex structure moduli as well. 
Fully contracted
combinations of Riemann tensors which do not depend on the complex 
structure are known \FreemanZH, and are the
generalizations of the Euler density ${\cal S}^{(1)}$   
$$
R_{r_1s_1}{}^{r_2s_2} R_{r_2s_2}{}^{r_3s_3} \cdots
R_{r_{3m}s_{3m}}{}^{r_1s_1} - 2^{3m-1} \, R_{r_1}{}^{r_2}{}_{s_1}{}^{s_2} R_{r_2}{}^{r_3}{}_{s_2}{}^{s_3} \cdots
R_{r_{3m}}{}^{r_1}{}_{s_{3m}}{}^{s_1} + {\rm Ricci}
$$ 
One thus reaches the  conclusion that ${\cal S}^{(m)}$ in \tendone\
are given precisely by these combinations. 

\noindent
{\bf Acknowledgments:}\hfill\break
We thank S. Ferrara for discussions and  encouraging remarks.  We also thank B. de Wit, V. Kaplunovsky, P. Kaste, P. Mayr,
K.S. Narain, S. Vandoren, A. Westerberg and especially T. Taylor for useful discussions. R.M. thanks the Theory
Division of CERN,  and S.T. thanks
the Centre de Physique Th{\'e}orique de l'{\'E}cole Polytechnique 
and the Phyiscs Department at University of Western Australia
for hospitality during the course of the work
and A. Font, K. F{\"o}rger, S. Kuzenko and S. Stieberger for helpful discussions. 
This work was supported in
part by the European Commission under the RTN contracts 
HPRN-CT-2000-00148/00131, the German-Israeli Foundation for
Scientific Research (GIF)
and the INTAS contracts 55-1-590 and 00-0334.

\appendix{A}{$R^4$ terms and corrections to the universal hypermultiplet}

This Appendix has two parts. First,  we collect some formulae needed
for reducing the effective action from ten or eleven
dimensions to four or five dimensions, respectively, on a Calabi-Yau
manifold. We then discuss the reduction of $R^4$ terms and their relation
to the corrections in the universal hypermultiplet.

\subsec{Useful formulae}
\item{$\triangleright$}{\bf Definitions:}

The connection is 
\eqn\eGamma{
\Gamma^M_{NP} = {1\over2} G^{MK} \left( \partial_N G_{PK} + \partial_P G_{NK}
- \partial_K G_{NP}\right)\ .
}
The Riemann tensor, defined as $[\nabla_M,\nabla_N]V_P=R_{MNP}{}^Q V_Q$, is
\eqn\eR{\eqalign{
R_{MNPQ} =& {1\over2} \left(\partial^2_{MQ} G_{NP} + \partial^2_{NP} G_{MQ}-
\partial^2_{MP} G_{NQ}- \partial^2_{NQ} G_{MP} \right)\cr
& + G_{KL}
\left(\Gamma^K_{MQ}\Gamma^L_{NP} -\Gamma^K_{MP}\Gamma^L_{NQ}  \right)
}}

\item{$\triangleright$}{\bf Weyl Rescaling:}

Under a Weyl rescaling of the metric $\bar G_{MN} = \Omega^2 \, G_{MN}$, the
Ricci scalar transforms as

\eqn\eWeyl{
\bar{\cal R} = \Omega^{-2} \, \left({\cal R} - 2 (D-1) \nabla^2 \ln\Omega -
(D-2)(D-1) (\nabla\ln\Omega)^2 \right)
}

\eqn\eWeylII{
\int d^Dx \sqrt{\bar g}\,\bar{\cal R}
=\int d^Dx \,\sqrt{g}\,\Omega^{D-2}\,\left[{\cal R} + (D-2) (D-1) (\nabla \ln\Omega)^2 \right]
}

\item{$\triangleright$} {\bf Compactification:}

We consider a background metric 
associated with the warped product space-time $\IR^D \times    {\cal M}_6$
\eqn\eG{
G_{MN}(x,y) = \pmatrix{ g_{\mu       \nu}(x) & 0 \cr 0 & v_6^{1\over 3}(x)\,
\gamma_{ij}(y)\cr},\qquad \int_{CY_3} \sqrt{\det(\gamma_{ij})}=1\ ,
}
where $l_s^6\, v_6$ is the volume of ${\cal M}_6$.
Since we are only interested in the universal multiplet part of the
reduction of string/M-theory on a Calabi-Yau threefold, 
we assume that the internal metric is independent of the coordinate $x\in
\IR^D$. The non-vanishing components of the connection are 
\eqn\eCon{\Gamma_{ij}^\mu    =-{1\over6}\gamma_{ij} v_6^{1\over3}\,\partial^\mu    \ln v_6\,,\quad\qquad 
\Gamma^i_{\mu    j}={1\over6}\delta^i_j\,\partial_\mu    \ln v_6
}
and $\Gamma_{\mu    \nu}^\rho,\, \Gamma_{ij}^k$. The latter, being the
connections constructed from  $g_{\mu    \nu}$ and $\gamma_{ij}$,
respectively, are independent of $v_6$. The components of the Riemann
tensor involving derivatives of the volume are
\eqn\eRcomponents{\eqalign{
R_{ijkl}(G)&=v_6^{1\over3}\hat R_{ijkl}(\gamma)-{1\over 36}\, v_6^{2\over3}\, (\partial_\mu   \ln v_6)^2
\, (\gamma_{ik} \gamma_{jl} - \gamma_{il} \gamma_{jk})\cr
R_{\mu    i \nu j}(G) &= -{1\over 36}\gamma_{ij}\ v_6^{1\over 3}\,
\left[6\nabla_\mu    \nabla_\nu \ln v_6 + \partial_\mu    \ln v_6\,
\partial_\nu \ln v_6 \right]\cr
}}
All other components are volume independent ($R_{\mu    \nu\rho\sigma}$), or
zero ($R_{\mu    ijk},\,R_{i\mu    \nu\rho}$  and $R_{ij\mu    \nu}$).
Other useful results  (needed for $d=6$) are
\eqn\eRicci{\eqalign{
R_{ij}(G)&=-{d\over 36}v_6^{1\over3}(\partial_\mu    \ln v_6)(\partial^\mu    \ln v_6)\gamma_{ij}
-{1\over 6}v_6^{1\over3}\nabla^\mu    \nabla_\mu    \ln v_6\,\gamma_{ij}
+\hat R_{ij}(\gamma)\cr
R_{\mu    \nu}(G)&=R_{\mu    \nu}(g)-{d\over6}\left(
\nabla_\mu    \nabla_\nu\ln v_6+{1\over6}(\partial_\mu    \ln v_6)(\partial_\nu\ln v_6)\right)
}}
from which  
\eqn\eRscalar{
{\cal R}(G)=-{1\over 36}d(d+1)(\partial_\mu    \ln v_6)(\partial^\mu    \ln v_6)-
{d\over3}\nabla^\mu    \nabla_\mu    \ln v_6+{\cal R}(g)
+v_6^{-{1\over3}}\hat{\cal R}(\gamma)\,.
}
Of course, in our case $\hat R_{ij}(\gamma)=0$.

\subsec{Reduction of the $R^4$ terms}

We study the reduction of the one-loop $R^4$ couplings in type~IIA or
M-theory in the background geometry specified by~\eG. This section is
similar to  the Appendix~A of~\refs{\BeckerNN} but differs in details; e.g.
we do not assume that the dilaton dependence is captured by the 
replacement $R_{MN}\to R_{MN}+2\nabla_M\nabla_N\phi_{10}$. 

\item{$\triangleright$} {\bf Correction to the volume kinetic term in four 
dimensions.}

We want to confirm the action~\eSv\ by considering the reduction of the
ten-dimensional $R^4$ term in the Calabi-Yau geometry specified by~\eG.
For this we rewrite the kinetic terms for ${\cal R}_{(4)}$, $v_6$ and $\phi_4$ in \eSv\ 
and \eString~as
\eqn\eSvv{\eqalign{
S_2=& \int d^4x \sqrt{g^\sigma}
\left\{(v_6\,e^{-2\phi_{10}}-\chi_1)\,{\cal R}_{(4)}
+{1\over6}[5\,v_6 e^{-2\phi_{10}}+
(1-3\mu    _1 )\chi_1+6\alpha]\, (d\ln v_6)^2\right.\cr
&\left.-[4\,v_6 e^{-2\phi_{10}}+(4\alpha-\mu    _1\,\chi_1)]\,d\ln v_6\,d\phi_{10}
+4[ v_6 e^{-2\phi_{10}}+\alpha] \, (d\phi_{10})^2 \right\}
}} 
The tree-level mixing between $\phi_{10}$ and $v_6$ arises from the 
reduction $\Box{7pt}_{10}\phi_{10}=\Box{7pt}_4\phi_{10}
+(\partial^\mu    \ln v_6)(\partial_\mu    \phi_{10})$.
The dilaton dependence is difficult to test, given the lack of
knowledge concerning its couplings in ten dimensions
\foot{E.g. the
kinetic term for the dilaton receives corrections from 5-point contributions
$R^3 (\nabla\phi_{10})^2$ which are 
not all captured~\refs{\PeetersUB} by the modified
connection scheme of Gross and Sloan~\refs{\GrossMW}.}.
Using the ten-dimensional 
${\cal O}(\alpha')$ on-shell condition $R_{MN}+2\nabla_M\nabla_N\phi_{10}=0$
the one-loop mixing may arise from the couplings 
$R_{MN}S^{MN}$ which we discuss below.

We nevertheless can test the coefficients for the volume kinetic term.
In type IIA, the one-loop $R^4$ terms are $t_8t_8 R^4 +{1\over4} E_8$ 
where $E_8=8!\times    \delta^{M_1\cdots M_8}_{N_1\cdots N_8}\,
R^{N_1N_2}_{M_1M_2}\times    \cdots R^{N_7N_8}_{M_7M_8}$. 
This can be expressed in terms of a basis $R_{4i},\, i=1,\dots,6$ and $A_7$
(using the notation of the Appendix~B.2 of \refs{\PeetersVW}) 
of seven scalars built from four Riemann tensors (since we are 
compactifying on a Calabi-Yau manifold, we do not need 
terms involving the Ricci tensor or Ricci scalar of the internal manifold)
\eqn\eJnot{\eqalign{
R_{41} &= {\rm tr}(R_{MN}R_{NP}R_{PQ}R_{QM})
\to-(S_2-{1\over4}S_1)\,{\cal A}~;\cr
R_{42} &= {\rm tr}(R_{MN}R_{NP}R_{MQ}R_{QP})
\to{S_1\over 4}\,{\cal A}~;\cr
R_{43} &= {\rm tr}(R_{MN}R_{PQ}) {\rm tr}(R_{MN}R_{PQ})
\to2 S_1\,{\cal A}~;\cr
R_{44} &= {\rm tr}(R_{MN}R_{MN}) {\rm tr}(R_{PQ}R_{PQ})\to 0~;\cr
R_{45} &= {\rm tr}(R_{MN}R_{NP}) {\rm tr}(R_{PQ}R_{QM})\to 0~;\cr
R_{46} &= {\rm tr}(R_{MN}R_{PQ}) {\rm tr}(R_{MP} R_{NQ})
\to S_1\,{\cal A}~;\cr
A_7 &= R^{PQRS} R_P{}^M{}_R{}^U R_M{}^V{}_Q{}^W R_{UVSW}\to(S_2-{1\over 4}S_1)\,{\cal
A}~, }}
where ${\cal A}={1\over9\, v_6}\,(\partial\ln v_6)^2$. 
We have also given the contributions of each scalar to the 
kinetic energy of the universal volume modulus $v_6$. 
$S_1= \hat R_{ijkl}\hat R^{klmn}\hat R_{mn}{}^{ij}$ and 
$S_2=\hat R_{ikjl}\hat R^{imjn}\hat R_m{}^k{}_n{}^l$ form a basis of scalars
built from three Riemann tensors. The indices are raised with the metric $\gamma^{ij}$. 
One can now show that \refs{\PeetersVW}\foot{
For a comparison with the expressions $J_o$
of~\refs{\BeckerNN,\FreemanZH}, one can use the
identity $R^r{}_{[mnp]}=0$ to find that
$ t_8t_8R^4 -{1\over 4} E_8 = 64\, J_o$. 
The six-dimensional Euler density is 
$$\eqalign{
E_6 &={1\over 3! \, (8\pi)^3} \, \epsilon_6\epsilon_6\, R^3
={1\over 12\, (2\pi)^3} \, (S_1 -2\, S_2+ {\rm Ricci})
}$$
}
\eqn\ettR{\eqalign{
t_8t_8 R^4 &= 192 R_{41} +384 R_{42} +24 R_{43} +12 R_{44} - 192 R_{45}-96 R_{46}\cr
{1\over 4} E_8 &= t_8t_8 R^4 +192 R_{46} -768 A_7 +64 {\cal R}{\cal S} -768
R_{MN} S^{MN} + \hbox{higher Ricci}\cr
}}
where ${\cal S}=S_{ij}\gamma^{ij}=S_1-2\, S_2+{}$Ricci is the
six-dimensional Euler density (defined analogously to $E_8$). 
${\cal R}\, {\cal S}$ contributes to the correction to  ${\cal R}_{(4)}$,
while $t_8t_8R^4$, $R_{46}$, $A_7$, ${\cal R}\, {\cal S}$ and
$R_{ij}S^{ij}$ contribute to the kinetic term of the volume. Because the
Ricci terms in~\ettR\ are affected by field redefinitions, we introduce
arbitrary coefficients
$c_{1,2}$ in front of $64\, {\cal R}\, {\cal S}$ and $-768\, R_{MN}S^{MN}$.
Using the components~~\eRcomponents\ before
integration over the zero modes, we get
\eqn\eRedc{
 64\, (S_1-2\, S_2)\, \left[c_1\, {\cal R}_{(4)} 
- {6+12\, c_1 - 7\, c_2\over6}\, (d\ln v_6)^2 \right]\,.
}
Comparison with \eSvv~ shows that neither of the values $ \mu    _1  =\pm 
2$ is matched by the naive choice $c_1=c_2=1$. Of course, there is no
reason that this naive choice should work. Having fixed the sign ambiguity
of $\mu    _1$ (which in principle can be done by a string computation), one
could use \eRedc\ to determine $c_{1,2}$ in the given background.


\item{$\triangleright$} {\bf Reduction of $C_3\wed t_8 R^4$}

The $\int\, C_3 \wed t_8 R^4$ coupling
\refs{\VafaWitten,\JDuffTLiuMinasian} does not contribute to the
two-derivative effective action in four dimensions. The indices should
split into five external $\mu    -$type indices and six internal $i-$type
indices. Plugging the expressions for the components of the Riemann
tensor~\eRcomponents, we find that
$C_{\mu      i j}$ and
$C_{ijk}$ only appear with higher derivative terms ${\cal
O}(\partial_\mu      ^4)$. Only $C_{\mu      _1\mu      _2\mu      _3}$ could be
associated with the two-derivate interaction originating from $\int \,
\epsilon^{\mu     _1\cdots \mu      _5} \, C_{\mu      _1\cdots \mu      _3}
t_8(R_{\mu      _4 i_1} R_{\mu      _5 i_2} R_{i_3i_4} R_{i_5i_6}) \,
\epsilon^{i_1\cdots i_6}$, but the antisymmetrization on the $\mu 
_{4,5}$-indices makes the term vanishing.
\appendix{B}{The Quaternionic K{\"a}hler geometry}

Following~\refs{\FerraraIK,\StromingerEB}, we study here loop
corrections to the quaternionic geometry. Our notation is the same with
that used in these papers.

Introduce the real vierbein 
$$
V=\pmatrix{ v_1 \cr v_2 \cr \bar v_2 \cr -\bar v_1}\ ,
$$
such that $ds^2={}^T V (\sigma_2\otimes \sigma_2) V=v_1\bar v_1+v_2\bar v_2$. 
We split the holonomy group $O(4) = Sp(1) \otimes Sp(1)$, as well as
the connection $\Omega := P + Q$ with
$$
P:= -{i\over 2} p\cdot     \sigma \otimes \Id_2;\quad Q:=-{i\over2}\Id_2
\otimes q\cdot     \sigma\, ,
$$
where $\sigma^{i=1,2,3}$ are the Pauli matrices
$$
\sigma^1=\pmatrix{0&1\cr 1&0}; \quad
\sigma^2=\pmatrix{0&-i\cr i&0};\quad \sigma^3=\pmatrix{1&0\cr 0&-1}
$$
The connections $P$ and $Q$ satisfy the tetrad postulate 
\eqn\eDefCon{
dV + (P+Q)\wed V=0\ .
}
For further reference, it is useful to remark that 
\eqn\ePQ{
P+Q =-{i\over2} \pmatrix{ p_3+q_3 &  p^- &  q^- & 0\cr
p^+ & -p_3+q_3 & 0 &  q^-\cr
q^+ & 0 & p_3 -q_3 &  p^-\cr
0 & q^+ & p^+ & -(p_3+q_3)}
}
where we have introduced $p^\pm     =p_1\pm     i p_2$ and $q^\pm     =q_1\pm     i
q_2$. The reality of the vierbein implies that $p_i$ and $q_i$ are real. 

We define the Sp(1)
curvatures as $\Omega^i = {i\over2} V^\dagger \wed (\sigma^i \otimes
\Id_2)V$, and the connection   $P$ satisfies the constraint~\refs{\FerraraIK,\BaggerTT}
\eqn\eSpCon{
dp^i + {1\over 2} \varepsilon^{ijk}\, p^j\wed p^k=\Omega^i\iff\cases{dp^+-i p^+\wed p^3=\Omega^1+i\Omega^2\cr
dp^- +i p^- \wed p^3 = \Omega^1-i\Omega^2\cr
dp^3 +{i\over2}  p^+ \wed p^- = \Omega^3\cr
}}
We study the perturbative deformations of the quaternionic geometry by
solving~\eDefCon\ and~\eSpCon\ order by order in $\exp(2\phi_4)$: $V=
\sum_n \exp(2n\phi_4) \, V_n$. The map between the sigma-model metric~\eCP\
and the dilaton can receive perturbative corrections as
$\rho^2=\exp(-2\phi_4)-2\hat\alpha\, \hat\chi + \cdots$.

\subsec{Tree-level solution}
The zero-th order solution is given by making the choice of the
vierbein
\eqn\eVo{
V_{\rm o}=\pmatrix{ u \cr v \cr \bar v \cr -\bar u}
}
The Sp(1) curvatures are 
\eqn\eSpo{\eqalign{
\Omega^1_{\rm o} &= i \, (\bar u\wed v +\bar v\wed u)\cr
\Omega^2_{\rm o} &= \bar u\wed v +u\wed \bar v\cr
\Omega^3_{\rm o} &= i\, (\bar u\wed u + v \wed \bar v)
}}
The $u$ and $v$ are the isometry invariant coordinates 
defined as in the main text (see
equation~\eQ\ for $\hat\chi=0$), and satisfy the relations
\eqn\euv{
du = {1\over 2} \, u\wed (v+\bar v);\quad dv= v\wed \bar v+ u\wed \bar u
}
Equation~\eDefCon\ is solved by 
\eqn\eZero{\eqalign{
p_{\rm o}^+&=2i \bar u; \quad p_{\rm o}^-=-2iu;\quad p^3_{\rm o} = {i\over2}\, (v-\bar
v)\cr
q_{\rm o}^+&=q_{\rm o}^+=0;\quad q^3_{\rm o}={3i\over2}\, (\bar v- v)
}}
\subsec{One-loop correction}
We consider now the one-loop  corrections  to the previous solution. The
loop counting parameter is $\exp(-\phi_4)=\rho$. The most general
expression for the one-loop correction to the vierbein which is invariant
under the three isometries and which leads  to a real metric is
\eqn\eVI{
V_1=e^{2\phi_4}\ \pmatrix{%
\alpha u+\bar\beta \bar u+\gamma v +\bar \delta \bar v \cr 
\alpha' u+\bar\beta' \bar u+\gamma' v +\bar \delta' \bar v \cr 
\bar\alpha' \bar u+\beta'  u+\bar\gamma' \bar v + \delta'  v \cr 
-(\bar\alpha \bar u+\beta  u+\bar\gamma \bar v +\delta  v )\cr 
}}
At the first order, the metric is given by
$ds^2= {}^T(V_{\rm o}+2 V_1)(\sigma_2\otimes\sigma_2) V_{\rm o}$.

\noindent
Because the perturbative $u$-coordinate contains the field-strength of the
RR-fields and the imaginary part of $v$-coordinate is the NS-axion, some
physical constraints have to be imposed on possible deformations of the
metric. Following arguments of \refs{\StromingerEB}, we find the
following restrictions on the parameters appearing in \eVI:
\item{i)} In order not to violate parity invariance, $vv$ and $\bar v \bar
v$ terms must appear in the  combination $vv+\bar v \bar v$.
This implies  that $\delta'$ is a real parameter.

\item{ii)} 
The real and imaginary parts of $u$ are the RR-field strengths. Since
string perturbation forbids any mixing between these fields (cf. footnote
4), $uu$ and $\bar u\bar u$ terms can only appear in the combination
$uu+\bar u\bar u$.  This forces $\beta$ to be a real parameter.
The same conclusion can also be reached using parity invariance, since 
a parity transformation acts as $u\leftrightarrow\bar u$. 

\item{iii)} Amplitudes of odd powers of RR-fields vanish in string
perturbation theory.   This implies that $\beta'+\delta=0$ and
$\gamma+\bar\alpha'=0$.

\medskip 
\noindent 
At the first order, the metric then reads
\eqn\eMetricI{\eqalign{
ds^2&= {}^T(V_{\rm o}+2 V_1)(\sigma_2\otimes\sigma_2) V_{\rm o}\cr
&= 2(u\bar u+v\bar
v)+ \beta (u^2+\bar u^2)+ \delta' (v^2+\bar v^2)+ (\alpha +\bar
\alpha)\, u\bar u+ (\gamma'+\bar\gamma') \, v\bar v\ .
}} 
Since the coefficients $\gamma$ and $\delta$ and the imaginary parts of $\alpha$
and $\gamma'$  do not appear in \eMetricI, we can set them to zero.
The vierbein then takes the form (we drop primes)
\eqn\eVii{
V_1=e^{2\phi_4}\, \pmatrix{%
\alpha u+\beta \bar u\cr
\gamma v + \delta \bar v \cr 
\gamma \bar v + \delta  v \cr 
-(\alpha \bar u+\beta  u )\cr 
}}
where all coefficients are real. 
It is not difficult to prove order by order that this is the most general 
parameterization of the perturbative corrections to the vierbein (affecting the
metric) for any $V_n$ which satisfies the
constraints i), ii) and iii).

\noindent
The first-order corrections to the Sp(1) connections are $\Omega^i_1 =  i\
V_{\rm o}^\dagger (\sigma^i\otimes \Id_2) V_1$:
\eqn\eSpi{\eqalign{
e^{-2\phi_4}\,\Omega^1_1 &= (\alpha+\gamma)\, \Omega^1_{\rm o}+i(\beta-\delta) \, 
(u\wed v-\bar u\wed \bar v)\cr
e^{-2\phi_4}\,\Omega^2_1 &= (\alpha+\gamma)\, \Omega^2_{\rm o}+(\beta+\delta) \, 
(u\wed v+\bar u\wed \bar v)\cr
e^{-2\phi_4}\,\Omega^3_1 &= 2\alpha\, \Omega^3_{\rm o}+ 2i(\alpha-\gamma)\, \bar v\wed v\cr
}}
The first-order variation of~\eSpCon\ is 
\eqn\eSpConI{
dp^i_1 + {1\over 2} \varepsilon^{ijk} \, (p^j_1 \wed p^k_{\rm o} + p^j_{\rm o}
\wed p^k_1) = \Omega^i_1
}
Using that $d \exp(2\phi_4)=-\exp(2\phi_4)\, (v+\bar v)$, these equations
imply
\eqn\ePI{
 p_1^+= 2i(\alpha-\delta)\, \bar u;\quad p_1^-=2i(\delta-\alpha)\, u;
\quad  p^3_1= i(\alpha-2\delta) \, (v-\bar v)
}
and the one-loop correction to the vierbein becomes 
\eqn\eViii{
V_1= e^{2\phi_4}\ \pmatrix{%
\alpha\, u\cr
2(\alpha-2\delta)\, v +\delta\,\bar v \cr 
2(\alpha-2\delta)\, \bar v +\delta\,  v \cr 
-\alpha\, \bar u\cr 
}\,.}
Solving the first-order linearization of~\eDefCon\ does not constrain
$\alpha$ and $\delta$.

\noindent We redefine $(\alpha,\delta)\to (\hat\alpha +3\hat\beta,\hat \beta)\times     
\hat\chi$, and rewrite the one-loop correction to the vierbein~\eViii\ as:
\eqn\eViv{
V_1= e^{2\phi_4}\hat\chi\ \pmatrix{%
(\hat\alpha+3\hat\beta)\, u\cr
2(\hat\alpha+\hat\beta)\, v +\hat\beta\,\bar v \cr 
2(\hat\alpha+\hat\beta)\, \bar v +\hat\beta\,  v \cr 
-(\hat\alpha+3\hat\beta)\, \bar u\cr 
}}
This two parameter deformation is obtained by linearization of the
Calderbank-Pedersen metric~\eQ, taking into account the one-loop correction to
the dilaton $\rho^2=\exp(-2\phi_4)-2 \hat\alpha\, \hat\chi$ and to the
potential $F(\eta,\rho)=\rho^{3/2} -  2\hat\beta\,\hat \chi\,
\rho^{-1/2}$. The analysis in \refs{\StromingerEB}~ only 
considers the special case $\hat\alpha\neq 0$ and $\hat\beta=0$. 
Note that $\hat\beta\neq0$ is in fact a redundant parameter and we have 
set $\hat\beta={1\over 2}$ in \eFI~and \eRP.

\subsec{Two-loop correction}

At two-loops, the most general parametrization of the modification of the
vierbein compatible with the requirements i), ii) and iii) and affecting the
metric is
\eqn\eVT{
V_2=e^{4\phi_4}\, \pmatrix{%
\tilde\alpha u+\tilde\beta \bar u\cr
\tilde\gamma v + \tilde\delta \bar v \cr 
\tilde\gamma \bar v + \tilde\delta  v \cr 
-(\tilde\alpha \bar u+\tilde\beta  u )\cr 
}}
where all parameters are real. Solving  
equation~\eSpCon\ at two-loop order shows, after tedious but 
straightforward calculation, that the vierbein is given by
\eqn\eVTii{
V_2=e^{4\phi_4}\, \pmatrix{%
\left({3\over2} \alpha^2-4\alpha\delta+3\delta^2+\tilde\delta\right)\,
u  \cr
\left(4\alpha^2-28\alpha\delta +{99\over2} \delta^2+3\tilde\delta\right)\, v+\tilde\delta \bar v \cr 
 \left(4\alpha^2-28\alpha\delta +{99\over2} \delta^2+3\tilde\delta \right)\,\bar v+\tilde\delta v\cr 
-\left({3\over2} \alpha^2-4\alpha\delta+3\delta^2+\tilde\delta\right)\, \bar u\cr 
}\, ,}
where $\alpha$ and $\delta$ already appeared in $V_1$ and 
$\tilde\delta$ is unrestricted. The two-loop order variation of~\ePQ\
does not restrict $\tilde\delta$ either. Switching to the variables
$(\alpha,\delta,\tilde \delta)\to (\hat\alpha+3\hat\beta, \hat\beta,
\hat\gamma)\times     \hat\chi$ the second-order modification of the vierbein becomes
\eqn\eVTiii{
V_2=e^{4\phi_4}\hat\chi\, \pmatrix{%
\left({3\over2}\hat\alpha^2+5\hat\alpha\hat\beta
+{9\over2}\hat\beta^2+\hat\gamma\right)\, u  \cr
\left(4\hat\alpha^2-4\hat\alpha\hat\beta+{3\over2}\hat\beta^2
+3\hat\gamma\right)\,v+\hat\gamma \bar v \cr 
\left(4\hat\alpha^2-4\hat\alpha\hat\beta+{3\over2}\hat\beta^2
+3\hat\gamma \right)\,\bar v + \hat\gamma\,   v \cr 
-\left({3\over2}\hat\alpha+5\hat\alpha\hat\beta +{9\over2}\hat\beta^2+\hat\gamma\right)\, \bar u\cr 
}}
This can be absorbed to a modification of the map between the dilaton and
$\rho$ at two-loops as 
\eqn\eDilRho{
\rho^2=e^{-2\phi_4}- 2\hat\alpha \, \hat\chi
-2(\hat\gamma-\hat\beta(\hat\beta+4\hat\alpha))\,\hat\chi^2 \,
e^{2\phi_4}\, ,
}
while keeping the potential $F(\eta,\rho)=\rho^{3/2} -2\hat\beta\,\hat\chi\,
\rho^{-1/2}$ unchanged.
\appendix{C}{Some details of the one-loop string computation}

\subsec{Dependence of the amplitude on the supercurrent insertions}

We consider~\eADef\ where all vertex operators are in their 
canonical ghost picture. To achieve saturation of the 
superconformal ghost charge we had to insert two floating holomorphic
picture changers~\refs{\VerlindeTX} $Y(z_{1,2})=\{Q_{BRST},2\xi(z_{1,2})\}$ 
and two floating anti-holomorphic
picture changers $\tilde Y(\bar w_{1,2})$. Using the BRST
invariance of the supercurrent and the vertex operators,
we derive that the only contribution arises from the action of the BRST-charge
on the measure of integration, which varies into a total derivative of the
moduli~\refs{\VerlindeTX},\foot{See as well~\refs{\PasquinucciIK} for an 
application to an amplitude closely related to the one studied in section~5.}
\eqn\eAVar{\eqalign{
\partial_{z_1} \eADef&= \sum_{s,\tilde s=1}^4
\int_{\cal F}{d^2\tau\over\tau_2}
\int\!d^{2}\!x\int\!d^{2}\!u\int\!d^{2}\!v\sum_{m=\tau,x,u,v}{\partial\over\partial m}\,Z(\tau,\bar\tau)~\cr
&
\langle\partial\xi(z_1)Y(z_2)\tilde Y(w_1)\tilde Y(w_2)\,V_{\rm NS}^{(-1,-1)}(x) V_{F}^{(-{1\over2},-{1\over2})}(u)
V_{\bar F}^{(-{1\over2},-{1\over2})}(v)\rangle_{s,\tilde s}\ .
}}
By ghost charge conservation only the ghost-charge ${}+2$ part of the picture
changer $\left.Y(z_2)\right|_{+2}=\{\oint b e^{2\phi-2\chi},2\xi(z_2)\}$ contributes to the
amplitude. 
This is important in the computation of section~5,  where we choose specific polarizations of the 
external states such that the tree level amplitude vanishes. For this choice 
\eAVar\ is also zero and we are thus free to place the 
picture changing operators at any convenient position and we do not have to 
worry about any contributions from the boundary of moduli space which 
might need regularization.   
Identical arguments and conclusions apply to the variation w.r.t. $z_2$ 
of the positions of the right-moving picture changers.

\subsec{Useful formulae}

Here, we collect some results needed for the calculation of the one-loop
amplitude  we describe in section~5.  

The bosonic correlation functions are reduced, via Wick's theorem, 
to two-point functions (and their derivatives): 
\eqn\eBosonic{
\langle X^\mu    (z)X^\nu(w)\rangle
=-\eta^{\mu    \nu}\ln|\chi(z-w)|^2\equiv -\eta^{\mu    \nu}G_B(z-w)}
with 
\eqn\eGB{
G_B(z-w)=\ln\left|{\theta_1(z-w)\over\theta_1'(0)}\right|^2
-{2\pi\over\tau_2}({\rm Im}(z-w))^2}
To be able to 
compute the internal ${\cal N}=(2,2)$ part, we consider  
compactifications on symmetric orbifolds. Our final conclusions should 
not depend on this choice. The SCFT is then realized via 
twisted complex free bosons $X^I$ and fermions $\psi^I$, 
$I=3,4,5$ with twists ${\bf h}_I=(h_I,g_I)$, where e.g. 
$X^I(z+1)=e^{2\pi i h_I}X^I(z)$ and 
$X^I(z+\tau)=e^{2\pi i g_I}X^I(z)$.  
Space-time supersymmetry requires $\sum_I \,h_I=\sum_I g_I=0$. 
We bosonize the fermions $\psi^I=\exp(i\phi^I)$. 
The internal spin fields 
in~\eVRR\ are expressed as $\Sigma(z,\bar z)=\Sigma(z)\tilde\Sigma(\bar z)$,
$\Sigma=\exp(i{\sqrt{3}\over2}H)$ with 
$\sqrt3\,H=\sum_{I}\,\phi^I$ the left U(1)-current of
the ${\cal N}=(2,2)$ superconformal theory with 
similar expressions for the right-movers 
$\tilde\psi^I$ and $\tilde Z(\bar z)$.
Then, for spin structure $s$ 
\eqn\eAint{
A^{\rm int}=\langle\Sigma_L(u)\bar\Sigma_L(v)\rangle_s = 
{\theta'{}^{3\over4}_1(0)\over\theta_1^{3\over 4}(u-v)}\prod_{I=3}^5\,
\theta_{s,{\bf h}_I}\left(\Half(u-v)\right)}
The normalization is fixed by matching the singularity as $u\to v$ 
and using $\langle {\bf 1}_{\rm int}\rangle=\prod_{I=3}^5\theta_{1,h_I}(0)$.   
${\bf 1}_{\rm int}$ is the unit operator in the internal sector.
To sum over the spin structures we will need to use the 
following Riemann identity which is valid for 
$h_1=g_1=0,\,h_2+h_3+h_4=g_2+g_3+g_4=0$:
\eqn\eRiemann{
\sum_{s=1}^4(-1)^s\prod_{i=1}^4\theta_{s,{\bf h}_i}(z_i)
=-2\prod_{i=1}^4\theta_{1,-{\bf h}_i}(v_i)}
where 
$$
v_i=A_{ij}z_j\qquad\quad{\rm with}\qquad\quad
A_{ij}={1\over2}\pmatrix{1&\phantom{-}1&\phantom{-}1&\phantom{-}1\cr
1&\phantom{-}1&-1&-1\cr1&-1&\phantom{-}1&-1\cr1&-1&-1&\phantom{-}1}\,.
$$

\vfill\supereject
\listrefs
\bye